\documentclass[fleqn,usenatbib]{mnras}

\usepackage[T1]{fontenc}

\DeclareRobustCommand{\VAN}[3]{#2}
\let\VANthebibliography\thebibliography
\def\thebibliography{\DeclareRobustCommand{\VAN}[3]{##3}\VANthebibliography}

\usepackage{graphicx}	
\usepackage{amsmath}	
\usepackage{amssymb}	
\usepackage{amsfonts}
\usepackage{caption}
\usepackage{subcaption}
\usepackage{xcolor}

\usepackage{scalerel,tikz}
\usetikzlibrary{svg.path}
\definecolor{orcidlogocol}{HTML}{A6CE39}
\tikzset{orcidlogo/.pic={\fill[orcidlogocol] svg{M256,128c0,70.7-57.3,128-128,128C57.3,256,0,198.7,0,128C0,57.3,57.3,0,128,0C198.7,0,256,57.3,256,128z}; \fill[white] svg{M86.3,186.2H70.9V79.1h15.4v48.4V186.2z} svg{M108.9,79.1h41.6c39.6,0,57,28.3,57,53.6c0,27.5-21.5,53.6-56.8,53.6h-41.8V79.1z M124.3,172.4h24.5c34.9,0,42.9-26.5,42.9-39.7c0-21.5-13.7-39.7-43.7-39.7h-23.7V172.4z} svg{M88.7,56.8c0,5.5-4.5,10.1-10.1,10.1c-5.6,0-10.1-4.6-10.1-10.1c0-5.6,4.5-10.1,10.1-10.1C84.2,46.7,88.7,51.3,88.7,56.8z};}}
\newcommand\orcidicon[1]{\href{https://orcid.org/#1}{\mbox{\scalerel*{
\begin{tikzpicture}[yscale=-1,transform shape]\pic{orcidlogo};
\end{tikzpicture}}{|}}}}

\newcommand{\mach}{\mathcal{M}}
\newcommand{\sigr}{\sigma_{\rho/\rho_0}}
\newcommand{\sigv}{\sigma_{v,\mathrm{3D}}}

\newcommand{\cc}{2.0 \times 10^{21}\,\mathrm{cm^{-2}}}
\newcommand{\pthreed}{P_\mathrm{3D}}
\newcommand{\ptwod}{P_\mathrm{2D}}
\newcommand{\cs}{c_\mathrm{s}}
\newcommand\hi{\mbox{\sc Hi}}
\newcommand{\bpk}{N_{\mathrm{bpk}}}
\newcommand{\brunt}{\mathcal{R}^{1/2}}

\title[Turbulence driving in the SMC]{A new method for spatially resolving the turbulence driving mixture in the ISM with application to the Small Magellanic Cloud}

\author[Gerrard et al.]{
Isabella~A.~Gerrard$^{\orcidicon{0000-0002-1995-6198}\,1}$\thanks{E-mail: isabella.gerrard@anu.edu.au},
Christoph~Federrath$^{\orcidicon{0000-0002-0706-2306}\,1,2}$\thanks{E-mail: christoph.federrath@anu.edu.au},
Nickolas~M.~Pingel$^{\orcidicon{0000-0001-9504-7386}\,3}$,
\newauthor\hspace{0.02cm}
Naomi~M.~McClure-Griffiths${\orcidicon{0000-0003-2730-957X}\,^1}$, Antoine~Marchal$^1$, Gilles~Joncas$^4 $, Susan E.~Clark$^{5,6}$, 
\newauthor\hspace{0.02cm}
Sne\v{z}ana~Stanimirovi\'{c}$^{\orcidicon{0000-0002-3418-7817}\,3}$, Min-Young~Lee$^7$, Jacco Th.~van~Loon$^8$, John~Dickey$^9$,
\newauthor\hspace{0.02cm}
Helga~D\'{e}nes$^{10}$, Yik~Ki~Ma$^{1}$, James~Dempsey$^{1,11}$, Callum~Lynn$^{1}$
\\
$^{1}$Research School of Astronomy and Astrophysics, Australian National University, Canberra, ACT 2611, Australia\\
$^{2}$Australian Research Council Centre of Excellence in All Sky Astrophysics (ASTRO3D), Canberra, ACT 2611, Australia\\
$^{3}$Department of Astronomy, University of Wisconsin-Madison, 475 North Charter Street, Madison, WI, 53706-15821, USA\\
$^{4}$D\'epartement de physique, de g\'enie physique et d'optique, Centre de recherche en astrophysique du Qu\'ebec, Universit\'e Laval, Qu\'ebec, G1V 0A6, Canada\\
$^{5}$Department of Physics, Stanford University, 382 Via Pueblo Mall, Stanford, CA 94305, USA\\
$^{6}$Kavli Institute for Particle Astrophysics \& Cosmology, P.O. Box 2450, Stanford University, Stanford, CA 94305, USA\\
$^{7}$Korea Astronomy and Space Science Institute, 776 Daedeok-daero, Yuseong-gu, Daejeon 34055, Republic of Korea\\
$^{8}$Lennard-Jones Laboratories, Keele University, ST5 5BG, UK\\
$^{9}$School of Natural Sciences, University of Tasmania, Private Bag 37, Hobart, TAS 7000, Australia\\
$^{10}$ School of Physical Sciences and Nanotechnology, Yachay Tech University, Hacienda San Jos\'e S/N, 100119, Urcuqu\'i, Ecuador\\
$^{11}$CSIRO Information Management and Technology, GPO Box 1700 Canberra, ACT 2601, Australia \\
}
\date{Accepted XXX. Received YYY; in original form ZZZ}

\pubyear{2023}

\begin{document}
\label{firstpage}
\pagerange{\pageref{firstpage}--\pageref{lastpage}}
\maketitle

\begin{abstract}
Turbulence plays a crucial role in shaping the structure of the interstellar medium. The ratio of the three-dimensional density contrast ($\sigr$) to the turbulent sonic Mach number ($\mach$) of an isothermal, compressible gas describes the ratio of solenoidal to compressive modes in the turbulent acceleration field of the gas, and is parameterised by the \emph{turbulence driving parameter}: $b=\sigr/\mach$. The turbulence driving parameter ranges from $b=1/3$ (purely solenoidal) to $b=1$ (purely compressive), with $b=0.38$ characterising the natural mixture (1/3~compressive, 2/3~solenoidal) of the two driving modes. Here we present a new method for recovering $\sigr$, $\mach$, and $b$, from observations on galactic scales, using a roving kernel to produce maps of these quantities from column density and centroid velocity maps. We apply our method to high-resolution $\hi$ emission observations of the Small Magellanic Cloud (SMC) from the GASKAP-HI survey. We find that the turbulence driving parameter varies between $b\sim 0.3$ and $b\sim 1.0$ within the main body of the SMC, but the median value converges to $b\sim0.51$, suggesting that the turbulence is overall driven more compressively ($b>0.38$). We observe no correlation between the $b$ parameter and $\hi$ or H$\alpha$ intensity, indicating that compressive driving of $\hi$ turbulence cannot be determined solely by observing $\hi$ or H$\alpha$ emission density, and that velocity information must also be considered. Further investigation is required to link our findings to potential driving mechanisms such as star-formation feedback, gravitational collapse, or cloud-cloud collisions.
\end{abstract}




\section{Introduction}
\label{sec:intro}
The interstellar medium (ISM) is turbulent and mostly composed of multi-phase hydrogen gas. The density distribution in cold, dense, self-gravitating molecular clouds embedded in the diffuse ISM is strongly influenced by the turbulence and magnetic field of the warm, diffuse medium from which they condense and are embedded in. The star formation rate and efficiency are functions of the density distribution of molecular clouds \citep{Mac-Low:2004aa,Elmegreen:2004aa, ScaloElmegreen2004,McKeeOstriker2007,HennebelleFalgarone2012,FederrathKlessen2012,PadoanEtAl2014}. Since the ISM is observed to be ubiquitously turbulent at all densities and temperatures, and since turbulence decays on short timescales \citep{StoneOstrikerGammie1998,MacLowEtAl1998}, ISM turbulence must be sustained by continuous energy injection into all phases of the ISM over a large range of scales.

Understanding the processes that drive the structure and evolution of the diffuse ISM, and the condensation of the warm neutral medium (WNM) to colder, denser gas phases from which stars are formed, is imperative to our continual quest for a coherent theory of ISM structure and star formation. Supernovae, stellar winds, protostellar outflows, radiative feedback, large-scale galactic dynamics and galaxy interactions will all impact the evolution of density and velocity structures of the diffuse ISM in different ways, but they do so primarily through their ability to precipitate and drive turbulence \citep{Elmegreen:2009tj,FederrathEtAl2017iaus}.

In order to characterise the turbulence in the ISM through comparison of observations, 3D simulations and analytic models, various statistical methods have been developed that are applicable to the information available in observational data \citep[see review by][]{Burkhart2021}. Measuring the spatial power spectrum is one commonly-used method, which reveals the energy injection and/or dissipation scale and the energy cascade in the turbulent ISM \citep[e.g.][]{Stanimirovic:1999aa, Stanimirovic:2001aa, KowalLazarian2007,HeyerEtAl2009,Chepurnov:2015aa,Nestingen-Palm:2017aa,Pingel:2018aa,Szotkowski:2019wa}. In addition, investigating the higher-order statistical moments, such as the skewness and kurtosis, or the probability distribution function (PDF) of column density is useful given the highly non-Gaussian behaviour of the density field of cold gas \citep{KowalLazarian2007, BurkhartEtAl2009, Burkhart:2010aa, Patra:2013aa, Bertram:2015aa, Maier:2017aa}.

In (magneto)hydrodynamic simulations of isothermal, supersonic gas it has been shown that the PDF of the gas density can be described by a log-normal function, meaning that the logarithm of the density follows a normal Gaussian distribution \citep{Vazquez-Semadeni:1994vq, PadoanJonesNordlund1997, Passot:1998wz, FederrathKlessenSchmidt2008,Hopkins2013PDF,SquireHopkins2017,BeattieEtAl2021}. It is the width (standard deviation) of this density PDF that describes the density fluctuations of the gas, and is a key parameter in theoretical models of star formation \citep{KrumholzMcKee2005, PadoanNordlund2011, HennebelleChabrier2011, FederrathKlessen2012,BurkhartMocz2019}. Furthermore, studies by \citet{PriceFederrathBrunt2011,KonstandinEtAl2012ApJ,MolinaEtAl2012,NolanFederrathSutherland2015,FederrathBanerjee2015,KainulainenFederrath2017} have shown that the width of the density PDF is proportional to the turbulent Mach number. Formally, this is described by
\begin{equation} \label{eq:b}
    \sigr = b\mach,
\end{equation}
where $\sigr$ is the three-dimensional (3D) standard deviation of density ($\rho$) scaled by the mean density ($\rho_0$), $\mach$ is the standard deviation of the (3D) turbulent velocity dispersion divided by the sound speed, and $b$ is a constant of proportionality known as the \emph{turbulence driving parameter} \citep{FederrathKlessenSchmidt2008,Federrath:2010wz}. This constant of proportionality can be used to quantify how turbulence is being driven in the gas: the \textit{acceleration field} that drives the turbulence can have purely solenoidal modes (divergence-free; $b=1/3$) or purely compressive modes (curl-free; $b=1$), or any combination of the two extremes. A value of $b=0.38$ is the natural mixture, which corresponds to 1/3 of the power in the driving field in compressive modes and 2/3 of the power in solenoidal modes \citep[see eq.~9 in][]{Federrath:2010wz}. The $b$ parameter has been derived in theoretical models, and tested in simulations \citep{FederrathKlessenSchmidt2008,Federrath:2010wz,PriceFederrathBrunt2011,MolinaEtAl2012,NolanFederrathSutherland2015,FederrathBanerjee2015,BeattieEtAl2021}. In addition, crucial methods have been developed to allow for a mapping between the intrinsically 3D properties of the gas ($\sigr$ and $\mach$), and the respective quantities that are accessible in observations, i.e., the column density and the intensity-weighted velocity centroid \citep{BruntFederrathPrice2010a,KainulainenFederrathHenning2014,Federrath:2016ac,Stewart:2022ut}. Thus, we can recover the turbulence driving parameter $b$ from observations, which allows us to learn more about what kind of physical processes are driving turbulence in those regions.

There have been several studies (primarily of molecular star-forming regions) in the Milky Way (MW) that have measured the $b$ parameter from observations of the variance in the density and velocity. In Taurus, \citet{Brunt2010} used $\mathrm{^{13}CO\,J=1-0}$ observations to derive $b=0.48^{+0.15}_{-0.11}$, which lies in the mixed--to--compressive regime, possibly a result of active star formation in that region. \citet{Ginsburg:2013ww} study a non-star-forming giant molecular cloud (GMC) towards W49A and find a lower limit for $b\gtrsim 0.4$, concluding that it is being driven compressively relative to the natural mixture case. The authors posit that because this GMC is representative of all GMCs in the solar neighbourhood, compressive driving may be a common feature of GMCs in general. \citet{Federrath:2016ac} find that the turbulence in a molecular cloud in the central molecular zone (CMZ), G0.250+0.016 (aka "The Brick"), is being driven solenoidally ($b \sim 0.22 \pm 0.12)$ due to the strong shearing motions in the CMZ, which may account for inefficient star formation in that region of the MW. In agreement with \citet{Ginsburg:2013ww}, \citet{KainulainenFederrath2017} find that the lower limit of $b$ for 15~solar neighbourhood clouds is $\gtrsim 0.4$ and rule out that any of these clouds can be dominated by solenoidal driving, although the authors point out that their density and velocity variance measurements do not always follow the standard relation described by Eq.~(\ref{eq:b}). \citet{Menon:2021vj} use ALMA observations of several CO isotopologues to measure the turbulence driving parameter in the "Pillars of Creation" in the Carina Nebula, and find that all the pillars are being compressively driven, with $b\sim 0.7 - 1$. The compressive driving in these star-forming regions could be a result of compression induced by photoionising radiation. There has also been one extra-galactic measurement of the $b$ parameter in the Papillon Nebula in the Large Magellanic Cloud (LMC) by \citet{Sharda:2022vg}, which is also a molecular star-forming region. They find that this region is being driven almost purely compressively, with $b\sim 0.9$, and speculate that this filamentary molecular cloud could have been formed by large-scale $\hi$ flows due to tidal interactions between the LMC and the Small Magellanic Cloud (SMC). \citet{Marchal:2021tz} derive a driving parameter for the warm atomic gas at high galactic latitudes in the MW using $\hi$ emission data, and found that the turbulence was driven mostly compressively in that region, with $b\sim0.68$. The scale on which all of these previous measurements of $b$ have been made is typically on the molecular cloud scale or smaller, and they all derive a single estimate for the turbulence driving parameter of those clouds/regions. This is largely because the resolution required to robustly estimate the density and velocity variance is available only for nearby regions of the ISM, and the star-forming molecular clouds are the most well studied. 

However, multi-phase atomic $\hi$ comprises the majority of the ISM, and understanding how turbulence driving affects its structure and evolution as it condenses into molecular hydrogen is crucial in understanding how molecular clouds form \citep{Gazol:2001wt, KoyamaInutsuka2002, AuditHennebelle2005, VazquezSemadeniEtAl2006, MandalFederrathKoertgen2020, Seta:2022ti}. Here we present a new method for mapping the turbulence driving parameter over large scales, and apply it to $\hi$ emission from the SMC, observed with the Australian Square Kilometre Array Pathfinder (ASKAP) \citep{Johnston:2008aa,Hotan:2021aa}. These data are unprecedented in both spectral and angular resolution (see Sec.~\ref{sec:obs}), and provide the ideal test bed for our new method of mapping the $b$ parameter, particularly because 21\,cm observations of the ISM provide a spatially coherent field in which we can probe the variations in density and velocity. Hitherto, no attempt has been made to spatially map the turbulence driving parameter over a large region, such as an entire galaxy that may contain an array of different physical turbulence driving mechanisms. Mapping the turbulence driving parameter requires high spatial and spectral resolution in order to accurately recover the density and velocity statistics, which makes $\hi$ particularly useful for this large-scale mapping technique.

The SMC is of particular interest because it has a rich dynamic history of interactions with the LMC and the MW. It is a disrupted, star-forming, low-metallicity, irregular dwarf galaxy \citep{Kallivayalil:2013aa}. There are, therefore, many physical mechanisms present by which to drive turbulence in the ISM across the SMC, providing a rich environment to test our method for mapping the $b$ parameter, and correlating our findings with known turbulence drivers.

The rest of this work is organised as follows. Sec.~\ref{sec:obs} introduced the data used in this work. Sec.~\ref{sec:meth} introduces our method for mapping the turbulence driving parameter, describing how we reconstruct the 3D density and velocity dispersions in large-scale observational datasets. Our results are presented in Sec.~\ref{sec:res}, and we discuss potential correlations of the $b$ parameter with $\hi$ and H$\alpha$ emission in Sec.~\ref{sec:correlations}. Sec.~\ref{sec:context} discusses our SMC measurements of the $b$ parameter in relation to other environments presented in the literature. We summarise our conclusions and pathways to future work in Sec.~\ref{sec:con}.

\section{Observations}\label{sec:obs}
\begin{figure*}
     \centering
     \includegraphics[width=\linewidth]{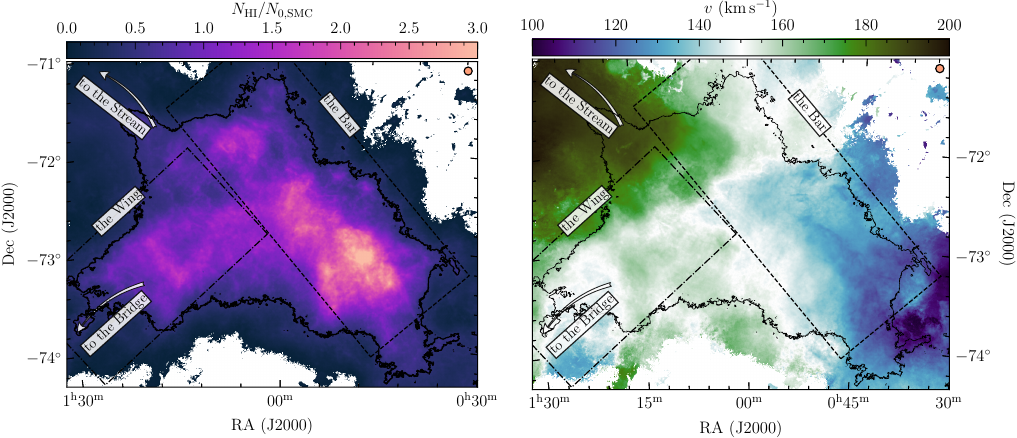}
     \caption{\emph{Left-hand panel}: Moment-0 map, which shows the integrated intensity. Because $\hi$ is (mostly) optically thin (See Appendix~\ref{sec:optical-depth} for further discussion on this topic), this quantity represents the column density $N_{\mathrm{HI}}$, which has been normalised by the mean of the emission inside the black contour, $N_{0, \mathrm{SMC}} = 4.5 \times 10^{21}$~cm$^{-2}$. The black contour (at $\cc$) denotes the first closed contour that encompasses the main body of the SMC. Following \citet{McClure-Griffiths:2018uf}, the overlays show the approximate location of the Bar (dashed box) and the Wing (dot-dashed box), and arrows indicate the directions towards the Magellanic Stream and the Magellanic Bridge linking to the LMC. The beam size is 30$''$, which is too small to show on this map. Each pixel in the map is 7$''$. The orange circle in the top right corner represents the full width at half maximum (FWHM) of the kernel used throughout the analysis in this work; it has a diameter of 10 instrument beams. \emph{Right-hand panel}: Same as the left-hand panel, but for the moment-1 map, which shows the intensity-weighted velocity centroid, $v$.}
     \label{fig:m0m1}
\end{figure*}

The observations of the SMC were obtained as part of the pilot phase of the atomic neutral hydrogen component of the Galactic ASKAP survey (GASKAP-HI), which aims to reveal the structure, kinematics, and thermodynamics of $\hi$ in the MW and Magellanic System at angular resolutions a factor of 3 to 30 times finer than existing 21-cm surveys of the Southern sky. The $\hi$ data cube\footnote{Download available at \url{https://doi.org/10.25919/www0-4p48}}, resulting from 20.9~hours of total integration with ASKAP, used for our analysis is the most sensitive (rms brightness temperature $\sigma_{T} = 1.1\,\mathrm{K}$ per $0.98\,\mathrm{km\,s}^{-1}$ spectral channel) and detailed view (30$''$ restoring beam or $\sim10\,\mathrm{pc}$ at the distance of the SMC) of the $\hi$ associated with the SMC made to-date. The details on the GASKAP-HI imaging pipeline, including the data validation, calibration, imaging, and combination with observations from the 64~m Parkes single dish telescope (Murriyang) to fill in the low spatial frequencies filtered out by ASKAP, are discussed in Sec.~3 of \citet{Pingel:2021vy}.

\subsection{Moment Maps}
Our analysis pipeline uses the moment-0 and moment-1 maps of some given position-position-velocity (PPV) cube as input, and processes them in such a way as to recover the 3D (volume) density dispersion and the Mach number, and thus the ratio of those two quantities -- $b$, as defined in Eq.~(\ref{eq:b}). The moment-0 (M0) is the integrated intensity, given by
\begin{equation} \label{eqn:M0}
\mathrm{M0} = \int T_b(v)\,dv, 
\end{equation}
where $T_b(v)$ is the brightness temperature (in Kelvin) in the channel, and $dv$ is the channel spacing.
The moment-1 (M1) is the intensity-weighted centroid velocity, and is given by
\begin{equation} \label{eqn:M1}
\mathrm{M1} = \frac{\int v\,T_b(v)\,dv}{\int T_b(v)\,dv} =\frac{\int v\,T_b(v)\,dv}{\mathrm{M0}}, 
\end{equation}
where $v$ is the velocity of each channel. We have $dv=0.98\,\mathrm{km\,s}^{-1}$, and integrate from $v=60.5\,\mathrm{km\,s}^{-1}$ to $v=235.0\,\mathrm{km\,s}^{-1}$.

By excluding any pixels below some multiple of the rms brightness temperature per channel prior to integration to create M0 and M1, we can be confident that the data we include in our analysis are robust and we do not introduce any spurious effects to our calculation of the turbulence driving parameter due to noisy pixels. For our data we choose a $10\sigma_T$ cut per channel,  (i.e., we only consider data with a signal-to-noise ratio $\geq10$). We discuss the effect of the choice of signal-to-noise threshold on our results in Appendix~\ref{app:snr}.

\subsubsection{Influence of the multi-phase $\hi$}
Eq.~(\ref{eq:b}) was derived for isothermal, hydrodynamic gas. $\hi$ emission does not originate from isothermal gas. The temperature structure of multi-phase neutral $\hi$ can be roughly categorised into cold neutral medium (CNM; $T \sim 100\, \mathrm{K}$), lukewarm neutral medium (LNM; $100\lesssim T/\mathrm{K}\lesssim6000$) and warm neutral medium (WNM; $T \gtrsim 6000\, \mathrm{K}$) \citep{Wolfire:1995aa, Wolfire:2003ug, Kalberla:2018tw}. The WNM is the most diffuse phase and has the largest volume-filling factor, while the CNM is expected to exist as smaller structures (such as clumps, sheets or filaments) dispersed throughout it \citep{Clark:2019aa}. As such, the emission from $\hi$ originates from a mixture of CNM, LNM and WNM. In order to apply Eq.~(\ref{eq:b}) to the $\hi$ emission data, we must make some careful assumptions about which phase dominates the moment-0 and moment-1 maps derived from the emission, which are the inputs to our analysis pipeline. 

The relative mass fractions of CNM, LNM and WNM will impact column density and the centroid velocity. For instance, narrow peaks in the spectrum due to the presence of CNM will shift the velocity centroid towards the centroid of the narrow peak, rather than measuring the centroid velocity of only the WNM. The strength of this effect, however, depends on the relative intensities of these CNM peaks with respect to the WNM component(s). The column density will also be affected by the presence of CNM, as the emission may be underestimated due to $\hi$ self-absorption (HISA) by the cold gas (discussed further in Appendix~\ref{sec:optical-depth}). Furthermore, the statistics associated with these quantities are also affected by the ratio of WNM/LNM to CNM. If we imagine that the $\hi$ structure is such that cold clumps/clouds of CNM are embedded in the diffuse WNM, then the dispersion of the centroid velocity will recover the intra-clump velocity dispersion, rather than the internal dispersion of the cold clumps themselves. Because the clumps are moving within the WNM, we can assume that this intra-clump velocity dispersion also traces the WNM velocity dispersion \citep{MohapatraEtAl2022, Kobayashi:2022vu}, which is what we primarily aim to measure. We also note that while the CNM may introduce a distortion in the individual line profiles, we only require a reasonably good estimate of the intensity-weighted average velocity along the LOS (i.e., the centroid velocity) for the velocity analysis (below in Sec.~\ref{sec:sigv}, which is not expected to be strongly affected by individual features in the line profiles. The effect that the CNM has on the column density and centroid velocity and associated statistics is a function of the relative CNM mass fraction. In the SMC, the CNM mass fraction is about 11\% \citep{Dempsey:2022vi}, meaning that the majority of the neutral $\hi$ is WNM/LNM. Furthermore, because the SMC is relatively metal-poor \citep{Russell:1992aa}, we expect that the portion of the gas that is not CNM is mostly comprised of WNM, as metal-line cooling will be less efficient in low-metallicity environments, as will photoelectric heating \citep{Bialy:2019aa}, and therefore the amount of LNM is likely much smaller than the WNM. At such low CNM percentages, the effects described above are relatively small, so we can assume that the emission data we use throughout this work is representative of the WNM, while acknowledging that the column density and centroid velocity maps are a product of a mixture of WNM, LNM and CNM.

\section{Methods}\label{sec:meth}
Here we present the analysis pipeline developed for producing a map of the turbulence driving parameter on galactic scales using the moment-0 and moment-1 maps of position-position-velocity (PPV) data as input. In this section we will describe each step in detail, but we begin with a brief summary of the steps in the pipeline:

The turbulence driving parameter $b$ in Eq.~(\ref{eq:b}) is constructed from the volume density dispersion $\sigr$ and the turbulent sonic Mach number $\mach$. To recover the volume density dispersion, we first compute the column (2D) density dispersion from the moment-0 map and use it to reconstruct the volume (3D) density dispersion via the Brunt method (Sec.~\ref{sec:sigrho}). The Mach number is a function of the 3D velocity dispersion and the sound speed, and we again start by finding the velocity centroid dispersion using the moment-1 map, and then extrapolating it to 3D (Sec.~\ref{sec:sigv}). To compute $\mach$, we then divide the 3D velocity dispersion by the sound speed of the gas (Sec.~\ref{sec:mach}). In order to isolate the true density and velocity fluctuations without contamination from non-turbulent motions due to bulk rotation and/or large-scale hierarchical density structure, we employ a gradient-correction method in our calculations of the density and velocity dispersion (Sec.~\ref{sec:gradsub}). This sequence of calculations is performed inside a Gaussian-weighted roving kernel (Sec.~\ref{sec:kern}), so as to build spatial maps of the 3D density variance, Mach number, and $b$ parameter. 

\subsection{Roving kernel}\label{sec:kern}
The density contrast and velocity fluctuations are scale-dependent quantities. Measuring the standard deviation of these quantities across the plane-of-the-sky requires some minimum number of spatial resolution elements. Our approach is to move a circular kernel across the input maps, pixel-by-pixel, calculate the dispersion within the kernel and assign that value to the central pixel, thus building up a map of the dispersion of the quantity from the input map (e.g., moment-0 or moment-1). The pixels of the resultant map are therefore not independent measurements and are correlated with one another on the scale of the diameter of the roving kernel.

In this study we choose a Gaussian kernel, defined by 
\begin{equation}
    \mathcal{K}(x,y) = \exp[-[(x-i)^2+(y-j)^2]/(2\sigma^2)],
\end{equation}
where $i$ and $j$ denote the central pixel of the kernel, $x$ and $y$ are the position of each pixel inside the kernel, and $\sigma$ is the full width at half maximum (FWHM) of the kernel divided by 2.355. We choose the number instrument beams per kernel FWHM to be 10, a quantity that is referred to throughout this text as $\bpk = D_{\mathrm{kernel}}/D_{\mathrm{beams}}$, where $D_{\mathrm{kernel}}$ is the FWHM of the kernel, and $D_{\mathrm{beams}}$ is the diameter of the instrument beam. This gives a sufficiently large number of independent data points of the centroid velocity and column density to reliably calculate the dispersion in those quantities \citep[see][for a discussion on the choice of the number of resolution elements per kernel; and a detailed study on the effects of the kernel size is presented below in Sec.~\ref{sec:kernelsize}]{Sharda:2018tk}. We apply the Gaussian kernel to a cut-out of a square region three times the FWHM of the kernel, which truncates the Gaussian. The size of the kernel is a flexible parameter in our pipeline and should be chosen as a function of the beam and pixel size of the input data. The kernel is also weighted by the primary beam response of the mosaic -- i.e., the sensitivity across the field of view -- so that pixels with lower sensitivity contribute less to the calculations of the dispersion. This weighting is an optional input to the pipeline, and must be the same shape as the moment-0 and moment-1 images, and should be a grid of values between zero and one, which is then multiplied with the Gaussian to give a grid of weights that can be used to weight the dispersion calculations.

\subsection{Gradient subtraction}\label{sec:gradsub}
In calculating both the column density dispersion and the centroid velocity dispersion, we apply the gradient-subtraction method described in \citet{Federrath:2016ac} and \citet{Stewart:2022ut}. We refer the reader to those works for a complete discussion on the details of the method, which we summarise in relation to this work here. Our aim is to compute the \emph{turbulent} dispersion of the centroid velocity and column density maps, and we therefore must eliminate any systematic gradients in motion or intensity which are caused by other factors besides turbulence, primarily the hierarchical column density structure (e.g., galaxies tend to be denser in their centres), and any bulk rotation of the gas. This method was previously applied in \citet{Federrath:2016ac, Sharda:2018tk, Sharda:2019wn, Menon:2021vj}, but only to the velocity field. Here we apply the method to both density and velocity as large-scale gradients are present in both quantities (seen in Fig.~\ref{fig:m0m1}). 

To do this, we fit a linear gradient to the field, on the scale of the kernel in which we perform our calculations, then subtract it from the original field. Formally, for some quantity $q(i,j)$ with pixel coordinates $i$ and $j$, we define a gradient
\begin{equation}
    \nabla_q = a + bi + cj, \label{eq:grad}
\end{equation}
where $a,b,c$ are fit parameters and $(i,j)$ is a point on the plane of the sky. After fitting for $a$, $b$, and $c$, we can compute the gradient-subtracted field,
\begin{equation}
    q'(i,j) = q(i,j) - \nabla_q. \label{eq:gradsubtract}
\end{equation}
For our purposes $q$ is either the normalised logarithmic column density $\log_{10}(N_{\mathrm{HI}}/N_0)$, where $N_0$ is the mean column density inside the kernel, or the intensity-weighted velocity centroid $v$.

\subsection{Density Dispersion}\label{sec:sigrho}
\begin{figure*}
    \includegraphics[width=\linewidth]{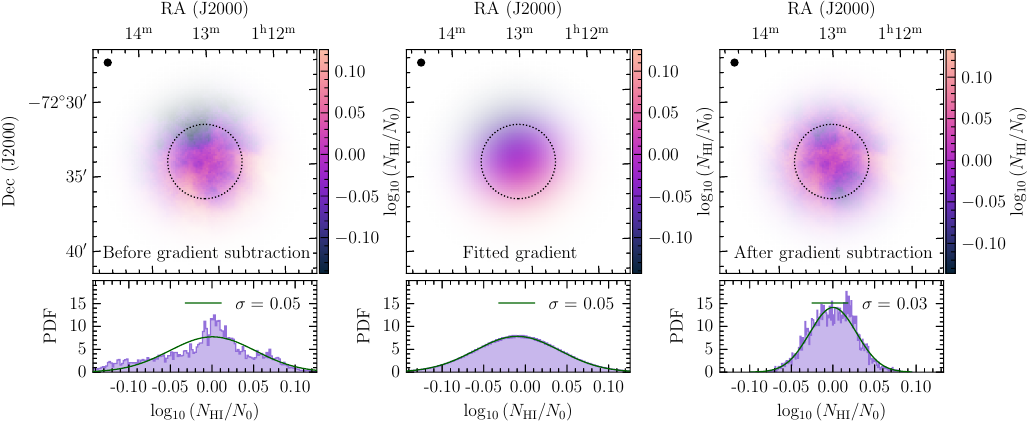}
    \caption{An example of a single kernel, illustrating the process of fitting a gradient to the column density map. In this example the column density ($N_\mathrm{HI}$) is normalised by the mean column density ($N_0$) in the kernel. The upper panels show the kernel-weighted maps (left: original column density map; middle: fitted gradient via Eq.~(\ref{eq:grad}); right: gradient-subtracted column density map via Eq.~(\ref{eq:gradsubtract}) ). We can see a distinct gradient in the original map, while the gradient-subtracted map shows a clearer density contrast. The black circles in each of the top panels shows the size of the instrument beam, while the dashed circles represent the kernel's FWHM (10 beams). The lower panels show the PDF in each case (shaded histogram), fitted with a Gaussian (solid line). The original data (left) has strong non-Gaussian components, as a result of the large-scale gradient. By contrast, the gradient-corrected data (right) is well-approximated by a Gaussian distribution in logarithmic column density, a universal feature of compressible turbulence.}
    \label{fig:densgrad}
\end{figure*}
To calculate the 2D column density dispersion, we take the moment-0 map as input (in this case the left-hand panel of Fig.~\ref{fig:m0m1}), and move the Gaussian kernel across it, computing the following steps for each pixel. We first normalise the column density by the mean column density inside the kernel (left-hand panel of Fig.~\ref{fig:densgrad}) and then fit a gradient to the normalised map, using Eq.~(\ref{eq:grad}) (middle panel of Fig.~\ref{fig:densgrad}). We then subtract the gradient from the original normalised map via Eq.~(\ref{eq:gradsubtract}), so that we are left with the turbulent density variations (right-hand panel of Fig.~\ref{fig:densgrad}). We then calculate the standard deviation of the resultant map, which gives the 2D column density dispersion, $\sigma_{N_{\mathrm{HI}}/N_0}$.

In Fig.~\ref{fig:densgrad}, we have fitted Gaussians to the PDFs in the lower panels. The raw data in the left-hand panel clearly does not follow the anticipated Gaussian distribution, but after applying the gradient subtraction we see in the right-hand panel that the resultant distribution well approximates a Gaussian, which is a hallmark of compressible isothermal turbulence\footnote{Note that this type of PDF is often referred to as the 'log-normal' density PDF, and indeed, this is what is shown by the fitted lines, i.e., a Gaussian fit to the logarithm of the column density \citep[e.g.,][]{Federrath:2013wt,Rathborne:2014wd,Federrath:2016ac,Khullar:2021wt}.}. 

\subsubsection{Conversion from 2D to 3D density dispersion}\label{sec:brunt}
\begin{figure}
    \centering
    \includegraphics[width=\linewidth]{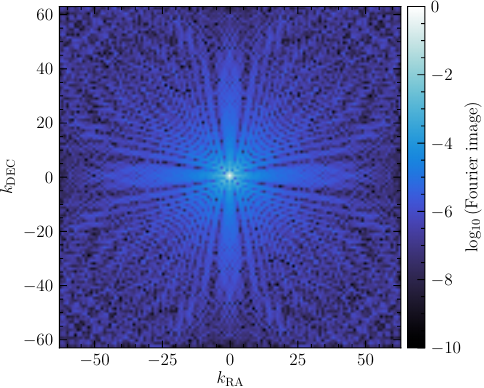}
    \caption{An example of the Fourier image of the gradient-subtracted column density shown in Fig.~\ref{fig:densgrad}. The wavenumber $k=1$ corresponds to $3\times D_{\mathrm{kernel}}$ in real space, and the subscripts `RA' and `DEC' denote the orientation in real space. Given the high level of point symmetry in this image, especially for small wavenumbers, which correspond to large length scales inside the kernel, the conversion from 2D to 3D density dispersion via the \citet{BruntFederrathPrice2010a} method is expected to introduce only a $\sim10\%$ uncertainty.}
    \label{fig:pspec}
\end{figure}
To convert the 2D density dispersion $\sigma_{N_{\mathrm{HI}}/N_0}$ to the 3D density dispersion, $\sigr$, we follow the method outlined in \cite{BruntFederrathPrice2010a}. This method is based on reconstructing the power spectrum of the volume density contrast from the power spectrum of the column density contrast, by measuring the column density power spectrum and extending its 2D power into 3D space.
The relation between the 2D density power spectrum $\ptwod(k)$ and the 3D density power spectrum $\pthreed(k)$ is given by \citep{Federrath:2013wt},
\begin{equation}
    \pthreed(k) = 2k \ptwod(k),
    \label{eqn:P3dP2d}
\end{equation}
where $k$ is the wave number. We exploit this relation to recover $\pthreed(k)$.

We first compute $\ptwod(k)$ of the gradient-subtracted column density,  ($N_{\mathrm{HI}}/N_0-1$ shown in Fig.~\ref{fig:pspec}), which immediately gives us $\pthreed(k)$ of the quantity $\rho/\rho_0$ as per Eq.~(\ref{eqn:P3dP2d}). The ratio of the sums over $k$-space of these two quantities gives the density dispersion ratio \citep{BruntFederrathPrice2010a},
\begin{equation}
    \mathcal{R}^{1/2} = \frac{\sigma_{N_{\mathrm{HI}}/N_0}}{\sigma_{\rho/\rho_0}} = \left(\frac{\sum_k \ptwod(k)}{\sum_k \pthreed(k)}\right)^{1/2},
    \label{eqn:brunt}
\end{equation}
known as the `Brunt Factor', which then allows us to recover the volume density dispersion, $\sigr$.

Eq.~(\ref{eqn:brunt}) relies on the assumption that the input field is isotropic in $k$-space, and that therefore $\ptwod(k)$ is also isotropic. This does not assume that the input field is isotropic in real space, only that its power spectrum is symmetric and isotropic, so we must check the Fourier image to make sure that this assumption holds, and we are not introducing further uncertainty in our measurements. The $\ptwod(k)$ image is shown in Fig.~\ref{fig:pspec}, where we can see that the power spectrum has good point symmetry. There are some angle-dependent features in the Fourier image. However, overall the distribution of power can be approximated as point symmetric around the $k=0$ mode (centre). The sampling of the power spectrum follows the spatial sampling of the data. Noise is accounted for through the SNR threshold applied (c.f., Sec.~\ref{sec:obs}). The effect of the beam size is accounted for in that it is primarily the low-$k$ modes (i.e., scales larger than the beam size) that contribute to the total power of the spectrum. We have also checked other regions (kernels) of the SMC and find qualitatively similar results. This implies that spherical symmetry is a good approximation for $\ptwod(k)$, such that the distribution of the density variations in real 2D space may be similarly distributed in 3D space.

\subsection{Velocity dispersion}\label{sec:sigv}
\begin{figure*}
    \includegraphics[width=\linewidth]{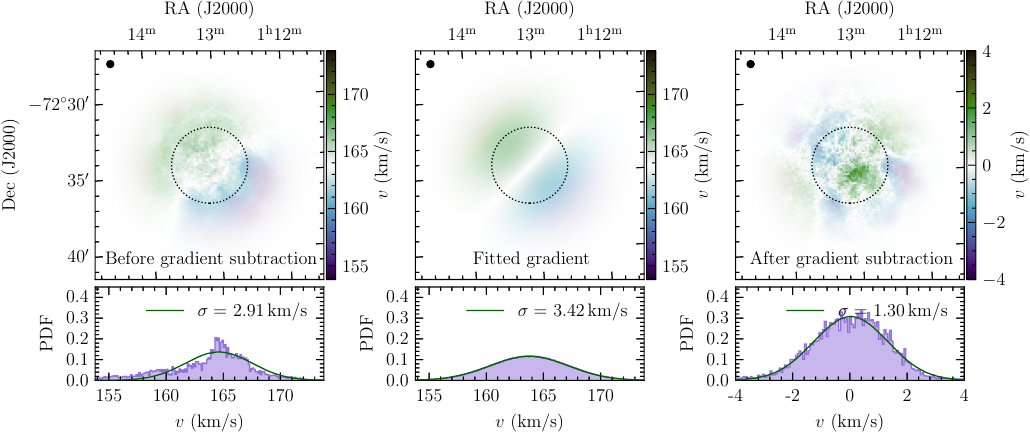}
    \caption{Same as Fig.~\ref{fig:densgrad}, but for the intensity-weighted velocity $v$ along the LOS (i.e., the moment-1). Similar to the column density, we find a significant gradient in the original data, leading to non-Gaussian components in the PDF of $v$. After gradient-correction, the PDF of $v$ follows a Gaussian distribution, which is a hallmark of turbulent flows. Thus, the gradient-subtraction method (c.f.~Sec.~\ref{sec:gradsub}) successfully filters out non-turbulent contributions and therefore isolates the turbulent velocity components in the data.}
    \label{fig:velgrad}
\end{figure*}
The 3D turbulent velocity dispersion is defined as
\begin{equation}
    \sigv = (\sigma_{v_x}^2 + \sigma_{v_y}^2 + \sigma_{v_z}^2)^{1/2},
\end{equation}
which cannot be measured in PPV space as we do not have access to the two velocity components in the plane of the sky. To recover $\sigv$ in a given kernel, we follow the methods developed by \cite{Stewart:2022ut}, who show that $\sigv$ can be recovered from PPV space using the standard deviation of the gradient-corrected moment-1 map together with a correction (1D/2D to 3D conversion) factor. As with our method for computing the density contrast, we apply a gradient correction to the moment-1 map that captures large-scale systematic motions (e.g., rotation) before computing the centroid velocity variance. First, we apply the kernel to the moment-1 map (right-hand panel of Fig.~\ref{fig:m0m1}). Next, we fit and subtract the gradient, and then take the standard deviation within the kernel to get the gradient-subtracted centroid dispersion $\sigma_{v,\mathrm{1D}}$. This process is illustrated in Fig.~\ref{fig:velgrad}.

Subtracting the large-scale gradients from the velocity map isolates the turbulent motions, however, the variance of the gradient-corrected map is not a true representation of the 3D turbulent velocity dispersion, because it does not take into account the line-of-sight dispersion. Therefore, we multiply by a correction factor\footnote{This correction factor is the mean of the $p_0$ values in lines~4--6 of Tab.~E1 of \citet{Stewart:2022ut}. We choose the gradient-subtracted statistics and choose the mean of those values, which are independent of the LOS orientation with respect to the rotation axis of the cloud/kernel region.} of $C_\mathrm{(c-grad)}^\mathrm{\,any}=3.3\pm0.5$, which was determined based on synthetic observations of 3D hydrodynamical simulations of rotating, turbulent clouds \citep{Stewart:2022ut} to convert the centroid velocity dispersion into a 3D turbulent velocity dispersion $\sigv$. The choice of using only the moment-1 map as opposed to using the moment-2 \citep[or the `parent' velocity dispersion, which is a combination of the moment-1 and moment-2; see][]{Stewart:2022ut} is discussed further in Sec.~\ref{sec:m1m2}.

\subsection{Mach number}\label{sec:mach}
The sonic Mach number ($\mach$) of the turbulent component of the velocity field is given by
\begin{equation}
    \mach = \frac{\sigv}{\cs}. 
    \label{eqn:mach}
\end{equation}
To construct a map of this quantity we need the velocity dispersion $\sigv$ (described above) and an estimate of the sound speed $\cs$. The last step in the pipeline is to divide $\sigv$ by the sound speed to produce the quantity $\mach$. Depending on available data, the sound speed could be input to our analysis pipeline as either a constant or as a spatially varying parameter (if spatially-resolved information about the temperature or sound speed is available). As we do not have access to spatially varying data for the sound speed, we choose a constant speed defined as
\begin{equation}
    \cs = \left(\frac{\gamma k_\mathrm{B} T}{\mu m_\mathrm{H}}\right)^{1/2},
    \label{eqn:cs}
\end{equation}
where $\gamma = 5/3$ is the adiabatic index, $k_\mathrm{B}$ is the Boltzmann constant, $T$ is the gas temperature, $\mu$ is the mean particle weight, and $m_\mathrm{H}$ is the mass of a hydrogen atom. Because we do not have a way to estimate the temperature of the combined $\hi$ phases present in our data, we will assume that the phase which dominates the temperature of the neutral gas is the WNM, and so we use its molecular weight of $\mu=1.4$ \citep{KauffmannEtAl2008}. In this study we adopt $T = (7.0 \pm 1.0) \times 10^3\, \mathrm{K}$, which gives a sound speed of $\cs = (8.3 \pm 0.6) \,\mathrm{km}\,\mathrm{s}^{-1}$. Estimates of the WNM temperature in the MW range from about 4000\,K to $10^4\, \mathrm{K}$, depending on the method used \citep{Wolfire:2003ug, Murray:2014aa, Murray:2018aa, Marchal:2021tz}. \citet{Bialy:2019aa} use simulations to investigate temperature and pressure structures in atomic neutral gas for varying metallicities. They find that the temperature structure changes for metallicity values smaller than $0.1\,Z_\odot$. The SMC is more metal poor than the MW ($Z \sim 0.2Z_\odot$, \citealt{Russell:1992aa}), and therefore choosing a temperature in line with solar metallicity values in accordance with \citet{Bialy:2019aa} seems a reasonable assumption. However, ultimately the temperature only enters with a square-root dependence in the sound speed, and therefore in the Mach number; hence, these assumptions do not introduce any major source of uncertainty in our calculations (discussed further in Sec.~\ref{sec:context} \& \ref{sec:cav}).

\section{Results} \label{sec:res}
In this section, we present and discuss the output of the analysis pipeline described in Sec.~\ref{sec:meth} as applied to the GASKAP-HI emission cube of the SMC. A summary of all the relevant physical parameters and measurements for the SMC is provided in Table~\ref{tab:vals}.
\begin{table*}
\centering
\def\arraystretch{1.15}
\caption{Summary of key quantities. The measured quantities are averaged within the closed contour, and therefore do not represent an actual global value for the entire SMC. The error bars for the quantities found in this work represent the range between the $16^{\mathrm{th}}$ and $84^{\mathrm{th}}$ percentile, or the variation around the median within the main contour, not actual uncertainties in those values. The quantities noted as `converged' are the best fit parameters found in Fig.~\ref{fig:convergence}, and their error bars are those produced by the fitting process.}
\begin{tabular}{ l l l l}
\hline
\hline
& \textbf{Symbol} & \textbf{Value} & \textbf{Comment/Reference} \\
\hline
\emph{\textbf{Constants}} & & & \\
Mean particle weight of WNM & $\mu$ & 1.4 & \citet{KauffmannEtAl2008} \\
Gas temperature & $T$ & $(7.0 \pm 1.0) \times 10^3$~K & Sec.~\ref{sec:mach}\\
Sound speed & $\cs$ & $8.3\pm 0.6\,\mathrm{km\,s^{-1}}$ & Derived from $T$ via Eq.~(\ref{eqn:cs})  \\
Velocity conversion factor & $C_\mathrm{(c-grad)}^\mathrm{\,any}$ & $3.3\pm0.5$ & \citet[][Tab.~E1, l.~4--6]{Stewart:2022ut} \\
\hline
\emph{\textbf{Measured \& Derived}} & & \\
Centroid velocity dispersion at $\bpk=10$ & $\sigma_{v, \mathrm{1D}}$ & $0.89^{+0.40}_{-0.24}$~km$\,\mathrm{s}^{-1}$ & This work \\
3D velocity dispersion at $\bpk=10$ & $\sigma_{v, \mathrm{3D}}$ & $2.91^{+1.32}_{-0.75}$~km$\,\mathrm{s}^{-1}$ & This work \\
Turbulent Mach number at $\bpk=10$ & $\mach = \sigma_{v, \mathrm{3D}}/\cs$ & $0.35^{+0.16}_{-0.09}$ & This work \\
Column density dispersion & $\sigma_{N_{\mathrm{HI}}/N_0}$ & $0.06^{+0.02}_{-0.01}$ & This work \\ 
Brunt factor (converged) & $\mathcal{R}^{1/2}$ & $0.31\pm0.01$ & This work \\
Volume density dispersion at $\bpk=10 $ & $\sigr = \sigma_{N_{\mathrm{HI}}/N_0}\,\mathcal{R}^{-1/2}$ & $0.18^{+0.06}_{-0.04}$ & This work \\
Driving parameter (converged) & $b=\sigr/\mach$ & $0.51\pm0.01$ & This work \\
\hline
\hline
\end{tabular}
\label{tab:vals}
\end{table*}

\subsection{Spatial distribution of volume density dispersion and turbulent Mach number}
\begin{figure*}
     \centering
     \includegraphics[width=\linewidth]{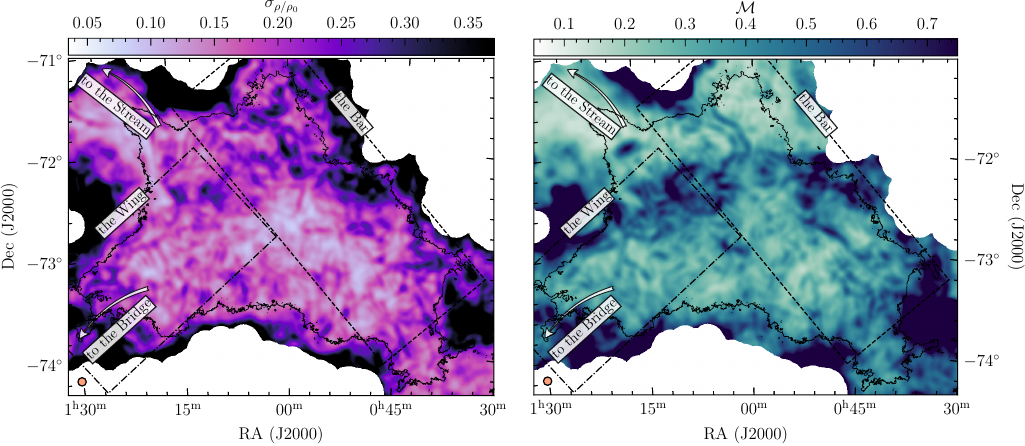}
     \caption{\emph{Left-hand panel}: 3D volume density dispersion $\sigr$. The black contour is the first closed contour at $\cc$. The orange circle shows the FWHM of the kernel with $\bpk = 10$ telescope beams per kernel. The white regions around the edges are a result of our SNR cut (see Sec.~\ref{sec:obs}). \emph{Right-hand panel}: Same as left-hand panel, but for the turbulent Mach number.}
     \label{fig:sigrM}
\end{figure*}
In the left-hand panel of Fig.~\ref{fig:sigrM} we present the volume density dispersion map, i.e., the plane-of-the-sky variations of $\sigr$, following the methods described in Sec.~\ref{sec:sigrho}. The quantity $\sigr$ measures the turbulent density variations in Eq.~(\ref{eq:b}). We see that the density dispersion varies by about an order of magnitude across the map. Within the analysis contour, the variations are about a factor of $\sim3$. We see a slight tendency of higher dispersion values towards the edges of the main contour (where the emission density begins to drop off), but overall the density dispersion does not show distinct regions of low or high values, with exception of the region at the top of the Wing and Bar regions. The relatively small variation and the lack of any large-scale global gradients in density dispersion within the main body of the SMC is a result of the gradient-subtraction method, which, although calculated on the kernel scale, successfully accounts for large-scale gradients (also) on the global scale. This is clear when visually comparing with the left-hand panel of Fig.~\ref{fig:m0m1}, where column density gradients towards the centre and bar of the SMC are visible. Given that these gradients would be expressed in the column density dispersion, and the volume density dispersion is directly related to that quantity (see Sec.~\ref{sec:brunt}), we would expect to see large-scale gradients in the volume density dispersion if the gradient subtraction method had not accounted for them. In summary, using the gradient-subtraction method, we have isolated the overall turbulent density fluctuations, which enter Eq.~(\ref{eq:b}).

The right-hand panel of Fig.~\ref{fig:sigrM} shows the map of the turbulent sonic Mach number, $\mach$, following the methods in Sec.~\ref{sec:mach}. The variations within the analysis contour are $\sim3$, similar to the variations of $\sigr$. The Mach number distribution is also relatively uniform within the analysis contour, with notable regions of high $\mach$ at the top of the Wing region, where it intersects with the Bar region, similar to $\sigr$. Comparing the $\mach$ map to the right-hand panel of Fig.~\ref{fig:m0m1} we can see that the gradient-subtraction method has successfully removed the large-scale rotation of the SMC. We find that the Mach numbers are distinctly subsonic within the main analysis contour, with typical values of \mbox{$\mach\sim0.1$--$0.7$}. Turbulent Mach numbers of this magnitude are expected for the WNM \citep{Burkhart:2010aa}, where the gas remains largely subsonic, while the CNM usually exhibits trans-sonic to mildly supersonic turbulence with \mbox{$\mach\sim1$--$2$} \citep{VazquezSemadeniEtAl2006, HennebelleAudit2007, HeitschEtAl2008}. \citet{Burkhart:2010aa} derived a spatial map of the sonic Mach number based on lower-resolution $\hi$ column density, and found that 90\% of the $\hi$ in the SMC was sub- or trans-sonic. While the primary beam of the data used in that work was much larger than the primary beam in the GASKAP-HI data, or the kernel FWHM we use in this work, the spatial distribution of the variations in the Mach number is in agreement. Higher Mach numbers towards the edges of the Wing, and surrounding the lower portion of the Bar are features in Fig.~\ref{fig:sigrM} of this work and Fig.~8 in \citet{Burkhart:2010aa}. The difference in absolute values of the Mach number is likely due to a difference in the methods used, as well as the difference in resolution of the two data sets. Increasing the kernel size (and thus mimicking the lower resolution in \citet{Burkhart:2010aa}) increases the Mach number range we recover in this work (discussed further in Sec.~\ref{sec:kernelsize} below), and pushes the values up into the trans-sonic/supersonic regime.

Both maps in Fig.~\ref{fig:sigrM} show some gradients at the edges of the fields, which are a result of lower emission density in those regions, and therefore lower SNR per-channel (below $\mathrm{SNR}=10$), as well as lower instrument sensitivity (40-80\% lower than inside the contour), which is why we have chosen to analyse only the data in the first closed contour.

\subsection{Spatial distribution of the turbulence driving parameter}\label{sec:moneyplot}
\begin{figure*}
    \includegraphics[width=0.95\linewidth]{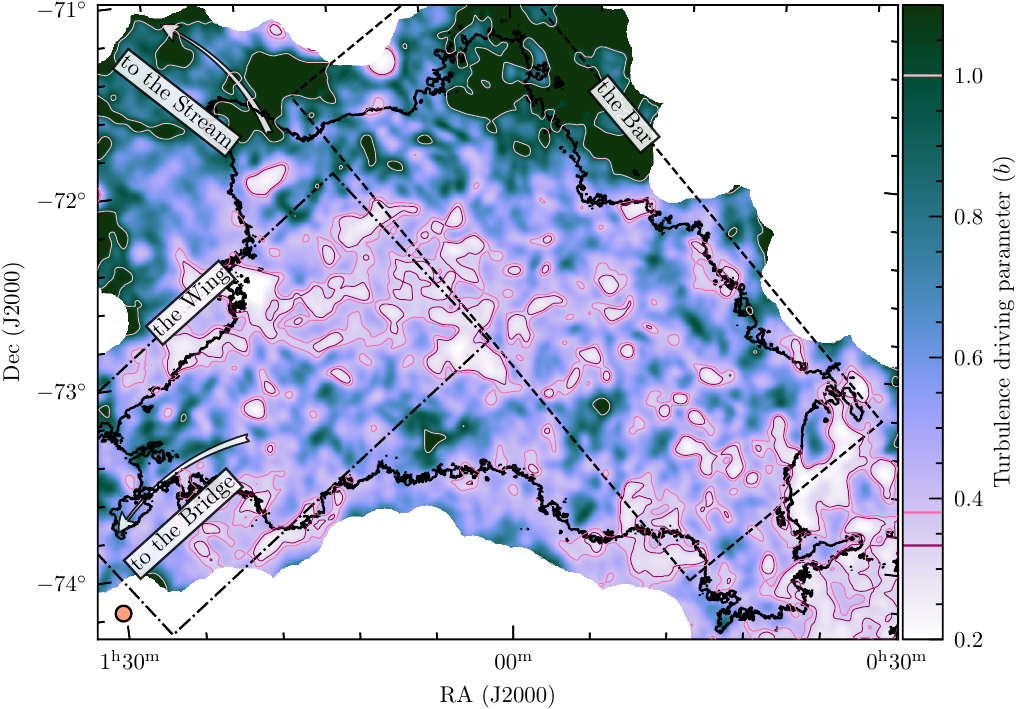}
    \caption{Map of the turbulence driving parameter, $b$, calculated via Eq.~(\ref{eq:b}), by combining the information in the maps shown in Fig.~\ref{fig:sigrM}. The main contour (black line) and kernel (shown in the bottom left corner) are the same as in Fig.~\ref{fig:sigrM}. The pink contours denote purely solenoidal driving (dark pink, $b\sim0.3$), naturally mixed driving (medium pink, $b\sim0.38$) and purely compressive driving (light pink, $b\sim1.0$). These lines are also shown on the colourbar. We see strong spatial variations in the turbulence driving parameter, ranging from purely solenoidal to purely compressive driving across the SMC.}
    \label{fig:bmap}
\end{figure*}
Combining the maps in Fig.~\ref{fig:sigrM} via Eq.~(\ref{eq:b}), we compute the turbulence driving parameter, $b$, for the whole SMC, which is shown in Fig.~\ref{fig:bmap}. It is immediately clear that there are spatial variations in $b$. Within the main contour, $b$ varies between $\sim0.3$ and~$1$, i.e., between purely solenoidal and purely compressive driving of the turbulence. Regions towards the Stream and the upper parts of the Bar seem to exhibit more compressive driving, while the central regions of the SMC tend towards more solenoidal and mixed ($1/3<b<0.38$) driving. The key result from this map is that we clearly find spatial variations in the turbulence driving mode. The exact cause of these variations is difficult to determine and requires detailed matching of the new $b$-parameter map obtained here, with information about potential physical drivers of turbulence \citep{Elmegreen:2009tj,FederrathEtAl2017iaus}, such as 1) feedback, i.e., from star formation/evolution activity, including supernovae, winds, jets/outflows, and radiation, or 2) dynamics of the Magellanic system, which includes accretion onto the SMC, large-scale flows inside the SMC, such as induced by rotation or shear, or tidal interactions with the hot Galactic halo causing ram pressure stripping. Disentangling and cross-matching all of these effects is challenging and will be subject of future work. However, we start investigating some of these potential correlations in Sec.~\ref{sec:correlations}, by studying the turbulence driving parameter as a function of the $\hi$ and H$\alpha$ emission in the SMC.

\subsubsection{Influence of the kernel size}\label{sec:kernelsize}
\begin{figure}
     \centering
     \includegraphics[width=0.9\linewidth]{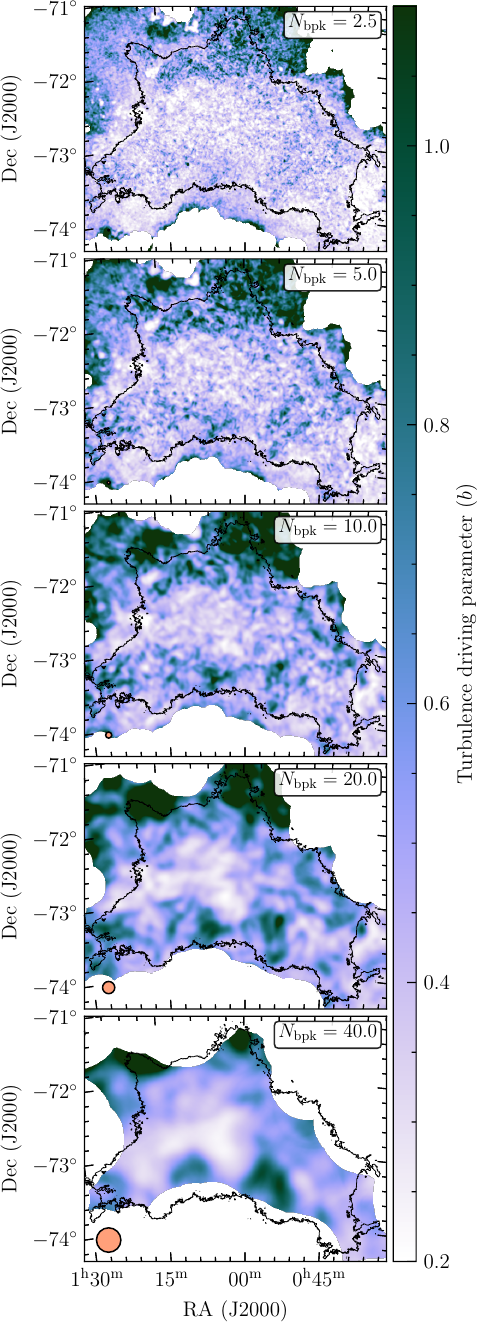}
     \caption{Comparison of the turbulence driving parameter map using different kernel sizes. From top to bottom, each panel uses a kernel with $\bpk = 2.5, 5.0, 10.0, 20.0, 40.0$. The orange circle in the bottom left-hand corner of each panel shows the kernel FWHM, and the black contour is the first closed contour as throughout this work.}
     \label{fig:rescomp}
\end{figure}
The 3D velocity dispersion and the volume density dispersion are scale-dependent quantities \citep{Kim:2005wz,KritsukEtAl2007,KowalLazarian2007,Federrath:2009wy,BeattieEtAl2019b,Federrath:2021wr}. In order to investigate the influence of the size of the analysis kernel, we compute the four main analysis quantities for 5~different kernel FWHMs, such that $\bpk =$ 2.5, 5.0, 10.0, 20.0, \& 40.0. The panels of Fig.~\ref{fig:rescomp} show the driving parameter for these 5~values of the kernel size, and highlight the different levels of spatial granularity resulting from each kernel size. All 5~kernel sizes we test are small compared to the size of the SMC. This should always be the case when applying this method to observations of entire galaxies, as the method is designed to probe the (small-scale) 3D turbulence properties.
\begin{figure*}
     \centering
     \includegraphics[width=\linewidth]{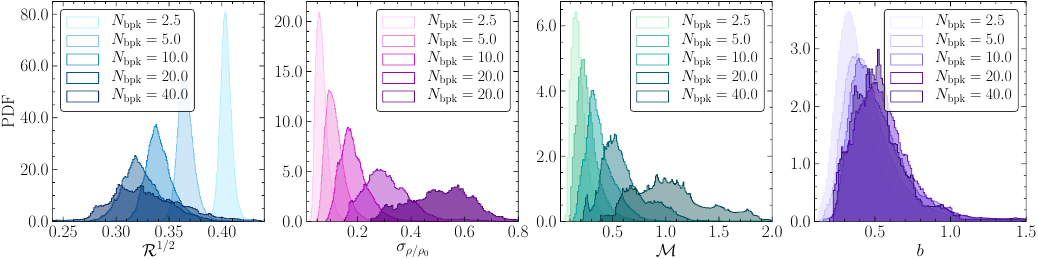}
     \caption{PDFs of the four main analysis quantities ($\brunt$ (blue panel), $\sigr$ (pink panel), $\mach$ (green panel) and $b$ (purple panel)) for five different kernel sizes. For the smallest kernel size ($\bpk = 2.5$), the peak of both $\sigr$ and $\mach$ are shifted towards smaller values, while the largest kernel size ($\bpk = 40.0$) shifts the peak of the distributions to higher values. On the other hand, the Brunt factor, $\brunt$, exhibits the opposite trend. While both $\sigr$ and $\mach$ depend on the kernel size, the ratio of the two, i.e., the turbulence driving parameter (Eq.~\ref{eq:b}) is independent of the kernel size for sufficiently large kernel size ($\bpk\gtrsim10$). At the distance of the SMC, the physical sizes of the kernels are approximately 25\,pc, 50\,pc, 100\,pc, 200\,pc and 400\,pc, respectively.}
     \label{fig:resPDF}
\end{figure*}

\begin{figure*}
     \centering
     \includegraphics[width=\linewidth]{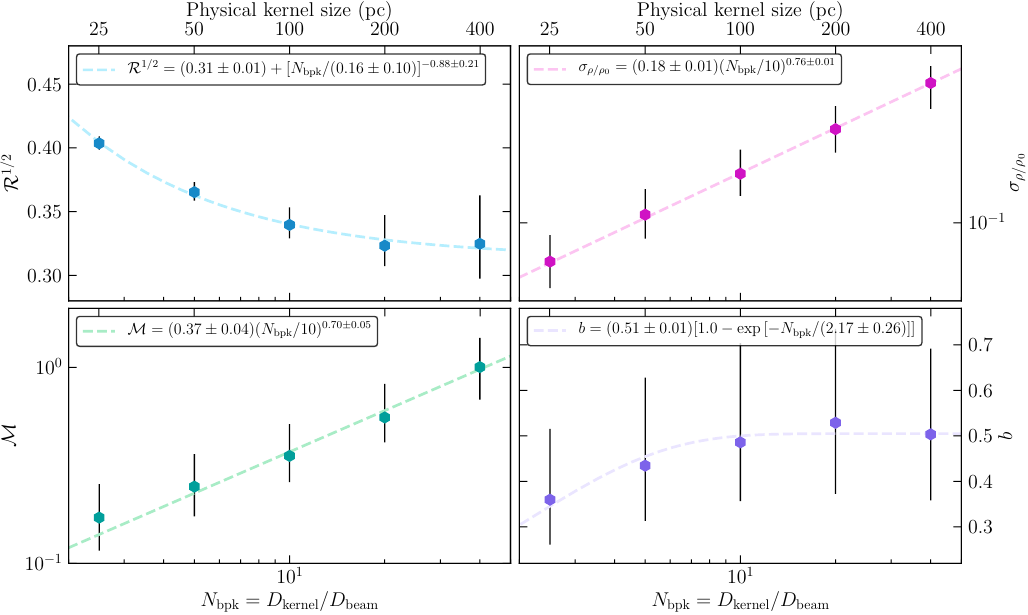}
     \caption{The median values of the four main analysis quantities as a function of $\bpk$, with fitted functions. \emph{Top left}: The Brunt factor $\brunt$, \emph{Top right}: The density variance $\sigr$. \emph{Bottom left}: The Mach number $\mach$. \emph{Bottom right}: The driving parameter $b$. The error bars on each data point represent the $16^\mathrm{th}$ and $84^\mathrm{th}$ percentiles, and were used in the fits, with the errors on the fit parameters displayed in the legend of each panel.}
     \label{fig:convergence}
\end{figure*}

In Fig.~\ref{fig:resPDF} we show PDFs of the 4~main quantities as a function of the kernel size (from left to right): the Brunt factor $\brunt$, the density dispersion $\sigr$, the Mach number $\mach$, and the driving parameter $b$. Fig.~\ref{fig:resPDF} demonstrates that the width and peak of the $\mach$ and $\sigr$ PDFs increase with increasing kernel size, and that the Brunt factor ($\brunt$) decreases with increasing kernel size. Because the width of the $\mach$ and $\sigr$ PDFs scale with kernel size in roughly the same way, the turbulence driving parameter $b$ is not a strong function of the kernel size (see right-hand panel of Fig.~\ref{fig:resPDF}). However, it is clear that the kernel size plays a role in the distribution and peak of all the analysis quantities, so in Fig.~\ref{fig:convergence} we plot the median value of the PDFs shown in Fig.~\ref{fig:resPDF} for each analysis quantity as a function of $\bpk$, to study the convergence behaviour of each of the relevant quantities.

Fig.~\ref{fig:convergence} shows how the median value of each analysis quantity behaves as a function of $\bpk$. The top two panels show the Brunt factor and the driving parameter, which both converge to a constant value as the kernel size increases. We find that the best fit for the Brunt factor (top left panel of Fig.~\ref{fig:convergence}) is
\begin{equation}
    \label{eq:bruntfit}
    \brunt = 0.31\pm0.01 + [\bpk/(0.16\pm0.10)]^{-0.88\pm0.21},
\end{equation}
such that in the limit of an infinitely large kernel ($\bpk\to\infty$) the Brunt factor converges to $\brunt = 0.31\pm0.01$. This value of roughly $0.3$ is consistent with previous findings \citep{Brunt2010,Ginsburg:2013ww,Federrath:2016ac,Menon:2021vj,Sharda:2022vg}.

The driving parameter (bottom right panel of Fig.~\ref{fig:convergence}) converges to a value of $b=0.51\pm0.01$, in the limit of an infinitely large kernel, which tends towards the compressive driving end of the scale ($b>0.38$). The functional form of the fit is given by
\begin{equation}
    \label{eq:bfit}
    b = (0.51\pm0.01)[1.0-\exp[-\bpk/(2.17\pm0.26)]].
\end{equation}
We take these two converged values to be the overall median Brunt factor and driving parameter in the SMC. The Mach number and volume density dispersion are not expected to converge, as they are scale-dependent quantities. We therefore fit power-law functions to each of these quantities. For the volume density dispersion we derive
\begin{equation}
    \label{eq:sigrfit}
    \sigr = (0.18\pm0.01)\,(\bpk/10)^{0.76\pm0.01}.
\end{equation}
and for the Mach number we find
\begin{equation}
    \label{eq:machfit}
    \mach = (0.37\pm0.04)\,(\bpk/10)^{0.70\pm0.05}.
\end{equation}
The exponents of these two power laws agree within the uncertainties, which shows that $\sigr$ and $\mach$ do indeed change in the same way with increasing kernel size, which can also be seen in Fig.~\ref{fig:resPDF}. Thus, $b$ converges to a constant value with increasing kernel size.

Because $\mach$ and $\sigr$ do not converge, we must necessarily choose one value of $\bpk$ when presenting the maps of these quantities and $b$. As previously outlined, we chose $\bpk = 10$, because it provided a sufficient level of granularity and had enough resolution elements per kernel to accurately recover the main analysis quantities (in particular the driving parameter $b$). We can see in Fig.~\ref{fig:convergence} that this choice of $\bpk$ recovers the converged value of $b$ within about 2.5\%. 

\begin{figure*}
     \centering
     \includegraphics[width=\linewidth]{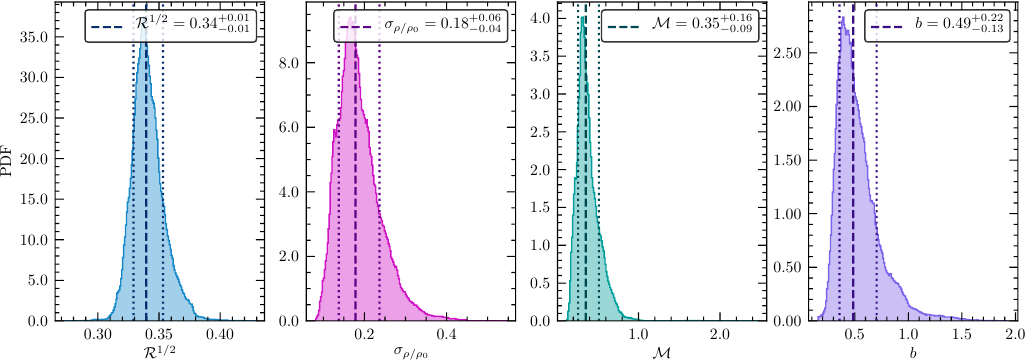}
     \caption{Same as Fig.~\ref{fig:resPDF} but for the $\bpk=10$ case. The solid lines in each panel represent the $50^{\mathrm{th}}$ percentile, while the dashed lines show the $16^{\mathrm{th}}$ and $84^{\mathrm{th}}$ percentiles, respectively. These values are also reported in the legend.}
     \label{fig:pdfs}
\end{figure*}

Fig.~\ref{fig:pdfs} shows the PDFs of $\brunt$, $\sigr$, $\mach$, and $b$ inside the main analysis contour for the standard kernel size of $\bpk=10$. We find $\brunt = 0.34\pm 0.01$, $\sigr = 0.18^{+0.06}_{-0.04}$ and $\mach = 0.35^{+0.16}_{-0.09}$. As expected for the WNM, the Mach numbers lie in the subsonic regime. For the turbulence driving parameter across the SMC, we find $b=0.49^{+0.22}_{-0.13}$. However, there are substantial variations of $b$ from region to region (as quantified by the $16^{\mathrm{th}}$ and $84^{\mathrm{th}}$ percentile range). Nevertheless, even the 1-sigma lower limit of $b$, i.e., the $16^{\mathrm{th}}$ percentile value is with $b_{16}=0.36$ still in the regime of a natural mixture of compressive and solenoidal driving. This is an interesting result as it may indicate that the $\hi$ gas is subject to predominantly compressive turbulence driving mechanisms in the SMC.


\section{Correlations between the turbulence driving parameter and the gas density and/or star formation activity}\label{sec:correlations}
\begin{figure*}
     \centering
     \includegraphics[width=\linewidth]{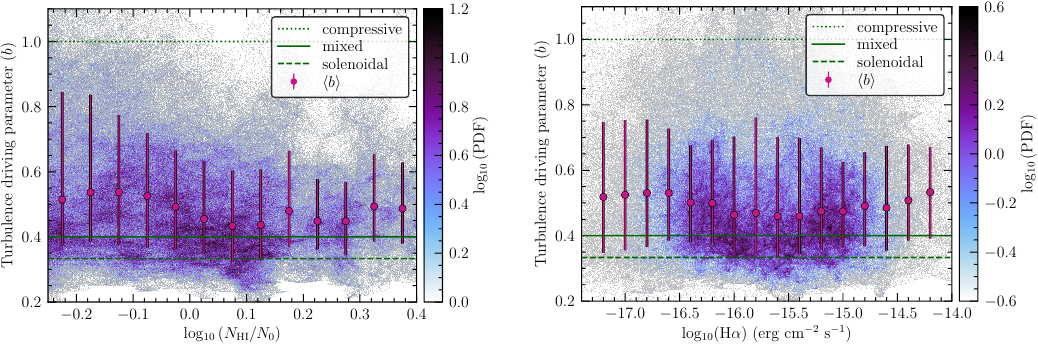}
     \caption{\emph{Left-hand panel}: Correlation between the $\hi$ emission and $b$. As in the left-hand panel of Fig.~\ref{fig:m0m1}, the mean column density is $N_0=4.5 \times 10^{21}$~cm$^{-2}$. The coloured data (with colourbar) shows the correlation PDF based on each spatial point in the main SMC contour, while the filled circles with error bars are the average $b$ values in bins across the $\hi$ range (with the position of the circle in the centre of the bin, and the extent of the error bars indicating the $16^{\mathrm{th}}$ to $84^{\mathrm{th}}$ percentile). The three horizontal lines show the theoretical limits for compressive ($b=1$, dotted), mixed ($b=0.38$, solid) and solenoidal ($b=1/3$, dashed) turbulence driving. \emph{Right-hand panel}: Same as left-hand panel, but for H$\alpha$ data from MCELS \citep{Smith:1999, Winkler:2015}.}
     \label{fig:HIHA}
\end{figure*}

In the left-hand panel of Fig.~\ref{fig:HIHA}, we investigate whether there is any correlation between $\hi$ intensity and the turbulence driving parameter $b$ within the main contour region of the SMC. We do not find evidence of a correlation between $b$ and $\hi$ intensity, instead the turbulence driving seems to be in the mixed--to--compressive ($b>0.4$) regime regardless of the $\hi$ emission density. This is expected since we found $b$ to be compressive overall in Fig.~\ref{fig:pdfs}. It also shows that $b$ is not simply a function of the column density, so one cannot simply assume that the turbulence is more compressively driven in regions of higher column density, in the case of $\hi$.
Similarly, we find that $b$ is slightly compressive across the entire range of H$\alpha$ \citep[MCELS][]{Smith:1999, Winkler:2015} intensity, as shown in the right-hand panel of Fig.~\ref{fig:HIHA}. To first order, the presence of H$\alpha$ signifies active, massive star formation, which may provide a feedback mechanism by which to drive turbulence compressively, through winds, expanding $\mbox{\sc Hii}$ regions \citep{Menon:2021vj}, outflows and ultimately supernovae \citep{Elmegreen:2009tj,FederrathEtAl2017iaus}. Compressive driving also positively influences star formation rates \citep{FederrathKlessen2012}, and so it is difficult to disentangle whether compressive driving where H$\alpha$ is present is causing star formation, or whether star formation is causing compressive driving. Because we do not find any correlation with increased H$\alpha$ intensity and more compressive driving, it is possible that the overall compressive nature of the driving in the $\hi$ gas in the SMC is not dominated by the star-formation activity.

\section{Comparison to other measurements of the turbulence driving parameter in different environments} \label{sec:context}
\begin{figure*}
  \includegraphics[width=\linewidth]{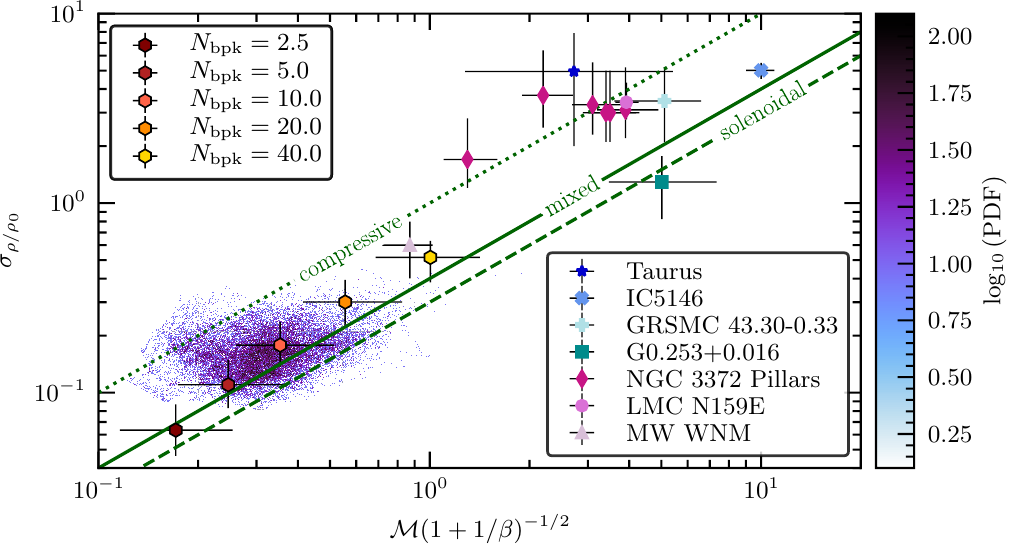}
  \caption{A summary of the available observational estimates for the turbulence driving parameter $b$ of several sources, contextualising our results for the SMC $\hi$ gas. The y-axis shows the 3D (volume) density dispersion ($\sigr$), and the x-axis shows the turbulent Mach number ($\mach$), including a factor involving plasma $\beta$ (ratio of thermal to magnetic pressure), as some of the literature values shown (Taurus and G0.253+0.016) have been calculated using the magnetised version of the density dispersion -- Mach number relation (see Sec.~\ref{sec:mag} for further discussion on this point). The three diagonal lines show the theoretical limits for compressive ($b=1.0$, dotted), mixed ($b=0.38$, solid) and solenoidal ($b=0.33$, dashed) driving of the turbulence \citep{Federrath:2010wz}. The SMC results of this work are shown in the lower left-hand corner, with the colourbar denoting the probability density of points in our SMC maps of $\sigr$ and $\mach$ for $\bpk=10$. The hexagons show the median values for the SMC quantities in the five different kernel sizes we investigated in Sec.~\ref{sec:kernelsize}, which correspond to roughly 25\,pc, 50\,pc, 100\,pc, 200\,pc and 400\,pc. The error bars on these points show the $16^{\mathrm{th}}$ to $84^{\mathrm{th}}$ percentile on each axis. For context we include a variety of sources from the literature: Taurus (dark blue star) \citep{Brunt2010}, which includes magnetic field estimates and revised Mach number estimations from \citet{Kainulainen:2013wh}, using $^{13}$CO line imaging observations; IC5146 (blue cross) \citep{PadoanJonesNordlund1997}, using $^{12}$CO and $^{13}$CO observations; GRSMC~43.30-0.33 (aqua plus) \citep{Ginsburg:2013ww}, observed in H$_{2}$CO absorption and $^{13}$CO emission; `The~Brick' (G0.253+0.016, teal square) \citep{Federrath:2016ac}, using HNCO observations; `The Pillars of Creation' (NGC~3372 pillars, magenta diamonds) \citep{Menon:2021vj}, from $^{12}$CO, $^{13}$CO and C$^{18}$O; `The Papillon Nebula' (LMC N159E, pink circle) \citep{Sharda:2022vg}, again in $^{12}$CO, $^{13}$CO and C$^{18}$O; and the WNM in the MW (lilac triangle) \citep{Marchal:2021tz} ($\hi$ observations).} 
  \label{fig:machsigr}
\end{figure*}

Fig.~\ref{fig:machsigr} concludes the discussion of our results by presenting a selection of other observational measurements of the density dispersion -- Mach number relation in regions of the MW, as well as one molecular star-forming region in the LMC. Here we can see that our data for the SMC lie between the lines denoting mixed and fully compressive driving, and that the points lie in the subsonic and low-density variance regime. With reference to our discussion in Sec.~\ref{sec:kernelsize}, however, it should be noted that the points we present on this figure (both our work and previous studies) are all a function of the scale (and selected region) on which they have been measured, which means that $\mach$ and $\sigr$ change with the kernel or region size observed. We show the median values for the five kernel sizes investigated in Sec.~\ref{sec:kernelsize}, corresponding to a physical size of 25\,pc, 50\,pc, 100\,pc, 200\,pc and 400\,pc. We can see that the smallest kernel (25\,pc) is squarely in the solenoidal regime (however, this value is not converged; see Fig.~\ref{fig:convergence}), and as the kernel size increases, compressive driving becomes more dominant (and leads to a converged value of $b$; c.f., Fig.~\ref{fig:convergence}).

Included in Fig.~\ref{fig:machsigr} are star-forming molecular clouds such as the Pillars~of~Creation \citep{Menon:2021vj}, Taurus \citep{Brunt2010, Kainulainen:2013wh}, IC5416 \citep{PadoanJonesNordlund1997}, and the Papillon~Nebula \citep{Sharda:2022vg}, which all exhibit supersonic, compressively-driven turbulence, likely driven by star-formation feedback. There are also two molecular clouds that are not star-forming; one which exhibits supersonic mixed/compressively-driven turbulence -- GRSMC~43.30-0.33 \citep{Ginsburg:2013ww}, and another, which is supersonic and solenoidally-driven -- `The~Brick' \citep{Federrath:2016ac}. The kind of turbulence driving in each of these molecular clouds depend on the specific physical mechanisms at play in each region, and while it is interesting to note that a variety of $b$ values are recovered and can be correlated to the environmental conditions \citep[for instance, strong shearing motions giving rise to predominantly solenoidal driving in the Brick][]{Federrath:2016ac}, it is difficult to do so for the present SMC measurements, as these contain contributions from the entire galaxy in the data shown in this plot. Correlating $b$ with environmental conditions will ultimately involve studying the spatial variation of $b$ as shown in the map of Fig.~\ref{fig:bmap}, and linking it to other measurements that provide information about the dynamics and potential physical drivers of turbulence, as discussed in the preceding sections.

The most direct comparison we can make with our data is the MW WNM datapoint \citep{Marchal:2021tz} (lilac triangle in Fig.~\ref{fig:machsigr}). \citet{Marchal:2021tz} found $\sigr = 0.6 \pm 0.2$ and $\mach = 0.87 \pm 0.15$, both somewhat higher than our measurements for the SMC. Because the density variation and the Mach number are scale-dependent quantities (see previous discussion; and Figs.~\ref{fig:rescomp}--\ref{fig:convergence}), we must first consider whether the spatial scale on which we measure our quantities is comparable to the scales on which the quantities in the MW were measured. We chose to use a kernel $\bpk=10$, which at the distance to the SMC is $\sim 100\,$pc. In \citet{Marchal:2021tz}, the spatial scale on which they measure $\sigr$ and $\mach$ is $\sim 63\,$pc. Referring to Fig.~\ref{fig:convergence}, we can see that had we used a kernel $\bpk=5$ ($\sim 50\,$pc), we would have derived even smaller median values for these two quantities. It is therefore likely that the difference between the measured SMC and MW values is not a result of the scale on which they were measured.

We have assumed that the WNM temperature in the SMC is $T = 7000\pm1000\,\mathrm{K}$, which is about 1000\,K higher than the temperature used in \citet{Marchal:2021tz}, but lower than the $10^4$\,K observed by \citet{Murray:2018aa}. There is a wide range of temperature estimates for the WNM in Galactic $\hi$, and we have chosen a temperature range in keeping with observations from \citet{Murray:2014aa}, following results from \citet{Bialy:2019aa} who show that at the metallicity of the SMC, the temperature structure should be similar to that of the MW. This influences our $\mach$ values to be about $10\%$ lower than if we had used the MW temperature estimate from \citet{Marchal:2021tz}. However, taking our lowest estimate for the temperature in the SMC and the highest estimate for the temperature in the MW, our median Mach number value still does not result in any overlap with the Mach number estimate for the WNM in the MW. We conclude that this cannot account completely for the difference between our Mach number values and those estimated in \citet{Marchal:2021tz}. It is likely that because we do not perform a phase separation as in \citet{Marchal:2021tz}, and assume that the emission is dominated by WNM, this is a contributing factor in our differing result, but still may not account for it entirely.

The difference between our median value for the volume density dispersion as compared to the reported MW value could be explained by significant variation in the depth of the SMC. If the SMC is highly extended along the line of sight, it is possible that we have underestimated the volume density dispersion via the \citet{BruntFederrathPrice2010a} method (Sec.~\ref{sec:brunt}). We discuss this issue further in Sec.~\ref{sec:depth}, but considering that we do not have a robust estimate for the extent of the SMC in the third dimension, we can only assume that our measured column density dispersion is a reasonable representation of the dispersion along the line of sight, and therefore $\sigr$ is truly smaller than in the MW WNM.

In summary, our analysis of the SMC WNM in comparison to the MW WNM region studied by \citet{Marchal:2021tz} may indicate that the values of density dispersion and Mach number are indeed physically smaller by a factor of \mbox{$\sim2$--$3$} in the SMC compared to the MW, but the average turbulence driving parameter of $b\sim 0.7$ for the MW and $b\sim 0.5$ for the SMC, indicates a similar dominance of compressive turbulence driving in the WNM of both the MW and the SMC.

\section{Caveats} \label{sec:cav}
\subsection{Magnetic fields} \label{sec:mag}
The ISM is ubiquitously magnetised, and the influence of magnetic pressure on the density dispersion -- Mach number relation has been derived theoretically and investigated in simulations \citep{PadoanNordlund2011,MolinaEtAl2012, Kortgen:2020uo} and observations \citep{Federrath:2016ac, KainulainenFederrath2017}. We have assumed the purely hydrodynamical relation in this work, as we are unable to map the magnetic field strength of the SMC in a way that would allow us to meaningfully incorporate it into our calculation of $b$. From \citet{MolinaEtAl2012}, we know that for cases in which the magnetic field strength is proportional to the square root of the density, $b$ is given by
\begin{equation} \label{eqn:molina2021}
    b = \sigr \mach^{-1} (1+\beta^{-1})^{1/2},
\end{equation}
where plasma beta, $\beta = P_{\mathrm{th}}/P_{\mathrm{mag}} = 2\cs^2/v_{\mathrm{A}}^2$, is the ratio of the thermal to magnetic pressure, and the Alfv\'en speed $v_{\mathrm{A}} = B_{\mathrm{turb}}/(4\pi\rho_0)^{1/2}$ for the turbulent component of the magnetic field, and $\rho_0$ is the mean volume density \citep[e.g.,][]{FederrathKlessen2012,Federrath2016jpp}. The best we can currently do is estimate a single value for $\beta$ in the SMC, and therefore quantify its bulk effect on our spread of $b$ values.

Our best estimate for the magnetic field strength across the SMC comes from \citet{Livingston:2022th}, who study Faraday rotation towards 80~sources across the SMC to estimate the line of sight magnetic field strength ($-0.3 \pm 0.1\,\mu$G) and the random component of the field ($5^{+3}_{-2}\,\mu$G). For the WNM the number density is $\sim 0.2 - 0.5\,\mathrm{cm^{-3}}$ \citep{Ferriere:2001vr}. This gives us an Alfv\'en speed between $v_{\mathrm{A}} \sim 5 - 30 \,\mathrm{km\,s^{-1}}$. Combining this we estimate $\beta \sim 0.1 - 2$. Using this in Eq.~(\ref{eqn:molina2021}), the factor $(1+\beta^{-1})^{1/2} \sim 1 - 3$, which means that the turbulence driving parameter could increase by up to a factor of three. \citet{Hassani:2022uz} estimate that $\beta < 1$ in the WIM and HIM, and although they make no estimate for $\beta$ in the WNM, it is in-keeping with our estimate. Using the magnetohydrodynamical version of the density dispersion -- Mach number relation does not provide a particularly useful constraint in this instance, but would generally increase our values of $b$, pushing our results towards the compressive end of the driving spectrum. However, given that we do not have a map of the random magnetic field strength across the SMC at this time, the hydrodynamic approach used in our analysis is robust enough to provide a lower limit of the turbulence driving parameter in the SMC.

\subsection{Calculating 3D velocity dispersion from the centroid velocity map} \label{sec:m1m2}
\citet{Stewart:2022ut} discuss three methods to determine the 3D turbulent velocity dispersion of a cloud. They find that the gradient-corrected parent velocity dispersion, the sum in quadrature of the gradient-corrected moment-1 and moment-2 maps, is the most robust way of recovering the 3D turbulent velocity dispersion of a cloud. We initially attempted to use this method, but were unable to reliably disentangle the various components along the LOS in a given pixel, causing an overestimation of the contribution of moment-2 to the sum. In future work we plan to explore new methods for decomposing multi-component spectra, at which time we will revisit this aspect of the pipeline. However, the unprecedented resolution of the PPV cube in both velocity and position allows us to use the correction factor found in \cite{Stewart:2022ut} ($C_\mathrm{(c-grad)}^\mathrm{\,any}=3.3\pm0.5$) to recover the LOS velocity fluctuations from the moment-1 map only, to within their quoted 10\% accuracy, which then gives us the 3D turbulent velocity dispersion.

\subsection{Depth of the SMC} \label{sec:depth}
The Magellanic system is highly dynamic and complex, and the 3D structure of the gas in the SMC is, for all intents and purposes, an unknown quantity. The robust gradient in the integrated velocity (see right-hand panel of Fig.~\ref{fig:m0m1}) has in the past been interpreted as evidence that the SMC has a disc-like structure \citep{Kerr:1954tg, Hindman:1963ux,Stanimirovic:2004ug, Di-Teodoro:2019ta}. However, the kinematics of gas-tracing stars, mapped using proper motions from \textit{Gaia}, are inconsistent with a rotational disc model \citep{Murray:2019aa}. 3D hydrodynamical simulations that attempt to reconstruct the dynamic history of the Magellanic system show that either one or two previous interactions between the SMC and LMC are required to consistently reproduce the Stream and the Bridge. \citep{Besla:2012tk, Diaz:2012vo, Lucchini:2021vd}.

It is likely that the SMC is not a rotating disc, but a torn and extended structure, with an estimated depth of \mbox{10--30~kpc} \citep{Tatton:2021aa}. This is not, however, a particular problem for our method. Because we use the \citet{BruntFederrathPrice2010a} method (Sec.~\ref{sec:brunt}) to recover only the volume density dispersion, rather than any estimate for the volume density itself, we do not need to directly account for the depth over which we integrate. The assumption behind this method is that the gas density dispersion along the line of sight is comparable to the density dispersion in the plane of the sky. We further find that the dispersion in the plane-of-the-sky is relatively isotropic (see Fig.~\ref{fig:pspec}), supporting the assumption in the \citet{BruntFederrathPrice2010a} method. We may be estimating a lower limit on the volume density fluctuations present in each kernel if the SMC is highly extended along the line of sight, given that integration over a large distance tends to wash out variance in the column density. However, the moment-1 map, used to quantify the turbulent velocity fluctuations in the plane-of-the-sky, is subject to similar smoothing along the LOS. Thus, both the moment-0 and the moment-1 maps are expected to be affected by LOS averaging in a similar way, and therefore, the ratio of the two (i.e., the turbulence driving parameter $b$; see Eq.~\ref{eq:b}) may be relatively robust against LOS averaging effects (similar to how it is independent of the kernel size on sufficiently large scales; c.f., Figs.~\ref{fig:resPDF} and~\ref{fig:convergence}). Without further exploration which is beyond the scope of this work and the available data, we cannot investigate this effect further at this time and therefore leave such an analysis to future work.

\subsection{Relative Uncertainties} \label{relerror}
As we have outlined in the above sections, there are several major assumptions made in this work, which likely dominate any errors associated with our results. However, under the assumptions we have made, we can quantify the relative error in our calculation of the turbulence driving parameter. We must account for the error introduced through the \citet{BruntFederrathPrice2010a} method ($\sim 10\%$) and the error associated with our estimate for the WNM temperature and therefore the sound speed ($\sim 10\%$). The error introduced by our conversion of the velocity variance via the \citet{Stewart:2022ut} method is $\sim 10\%$, which is similar to the range of applicable values, so the choice of conversion factor does not have a significant effect on our results. Combined, this gives a relative uncertainty of $\sim 20\%$ associated with our $b$ values. Referring to Fig.~\ref{fig:pdfs}, we can see that this error is smaller than the variance of $b$ itself, and as such, we are primarily measuring the spatial variations of the turbulence driving parameter across the SMC, rather than variations introduced due to noise or uncertainties in the method.

\section{Conclusions} \label{sec:con}
In this work we presented a new generalised method for creating a map of the turbulence driving parameter in the ISM of galaxies using column density and centroid velocity information. Our method can be applied on scales from molecular clouds to entire galaxies, provided there are enough spatially coherent resolution elements available (Sec.~\ref{sec:kernelsize}), and the data are sufficiently sensitive to provide reliable moment-0 and moment-1 maps. We use a Gaussian-weighted roving kernel (Sec.~\ref{sec:kern}) to recover the volume density dispersion and turbulent Mach number, and construct maps of these quantities (Fig.~\ref{fig:sigrM}) which we then use to create a map of the turbulence driving parameter (Fig.~\ref{fig:bmap}) via Eq.~(\ref{eq:b}). In summary, the main steps of the pipeline performed in each instance of the roving kernel are:
\begin{enumerate}
    \item To isolate only turbulent density variations, fit and subtract a linear gradient from the normalised column density map (see Fig.~\ref{fig:densgrad}).
    \item Via the "Brunt Method" (Sec.~\ref{sec:brunt}), use the 2D power spectrum of the gradient-corrected column density to construct the 3D power power spectrum, the ratio of which is the Brunt factor, $\brunt$.
    \item Using $\brunt$, reconstruct the volume density dispersion from the column density dispersion (see Fig.~\ref{fig:sigrM}).
    \item Fit and subtract a linear gradient from the centroid velocity map, thus isolating only turbulent velocity fluctuations (see Fig.~\ref{fig:velgrad}).
    \item Find the dispersion of the gradient-corrected centroid velocity, and convert to 3D velocity dispersion via the "Stewart" method (see Sec.~\ref{sec:sigv}.
    \item Divide the 3D velocity dispersion by the sound speed to recover the turbulent sonic Mach number, $\mach$ (see Fig.~\ref{fig:sigrM}).
    \item Finally, divide the volume density dispersion map by the Mach number map to create a map of the turbulence driving parameter, $b$, via Eq.~\ref{eq:b} (see Fig.~\ref{fig:bmap}).
\end{enumerate}

We have applied this method to high-resolution GASKAP-HI observations of 21\,cm emission from the SMC (Sec.~\ref{sec:obs}). We find that spatial variations of the turbulence driving parameter span the entire driving range from purely solenoidal to purely compressive driving, with some regions exhibiting very compressive driving ($b\sim1$, e.g., towards the Bridge and the Stream), while other regions are driven more solenoidally ($b\sim 1/3$, e.g., in the centre). We find that the driving parameter is a weak function of the scale on which it is measured (see Fig.~\ref{fig:convergence}), but that it converges on kernel scales $\gtrsim100\,\mathrm{pc}$) to a constant value of $b\sim 0.5$, which is towards the compressive end of the spectrum (i.e., $b>0.38$, which defines the natural driving mixture), and with 16th to 84th percentile variations in $b$ between $\sim 0.3$ and $\sim 0.6$ across the SMC. In the context of other measurements of the turbulence driving parameter in the WNM, specifically \citet{Marchal:2021tz}, we find that while both the volume density dispersion and the Mach number are significantly lower than MW values, $b$ is similarly compressive overall ($\sim 0.7$ in the MW). We do not find evidence for a correlation of $b$ with either $\hi$ or H$\alpha$ emission intensity. In future work we will delve deeper into specific regions of the SMC and correlate variations in the $b$ parameter with known physical turbulence driving mechanisms.

\section*{Acknowledgments}
The authors acknowledge Interstellar Institute's program "With Two Eyes" and the Paris-Saclay University's Institut Pascal for hosting discussions that nourished the development of the ideas behind this work.
I.A.G.~would like to thank the Australian Government and the financial support provided by the Australian Postgraduate Award. 
C.F.~acknowledges funding by the Australian Research Council (Future Fellowship FT180100495 and Discovery Projects DP230102280), and the Australia-Germany Joint Research Cooperation Scheme (UA-DAAD). C.F.~further acknowledges high-performance computing resources provided by the Leibniz Rechenzentrum and the Gauss Centre for Supercomputing (grants~pr32lo, pr48pi and GCS Large-scale project~10391), the Australian National Computational Infrastructure (grant~ek9) in the framework of the National Computational Merit Allocation Scheme and the ANU Merit Allocation Scheme, through which the data analyses presented in this paper were performed. S.S. and N.M.P. acknowledge support provided by the University of Wisconsin -- Madison Office of the Vice Chancellor for Research and Graduate Education with funding from the Wisconsin Alumni Research Foundation, and the NSF Award AST-2108370.

\section*{Data Availability}
The code used in the analysis is freely available on \href{https://github.com/igerrard/b_finder}{GitHub}. The GASKAP-HI emission cube used in this study is freely available to download at \url{https://doi.org/10.25919/www0-4p48}.

\bibliographystyle{mnras}
\bibliography{gerrard,federrath} 

\appendix
\section{Influence of the SNR} \label{app:snr}
\begin{figure*}
     \centering
     \includegraphics[width=\linewidth]{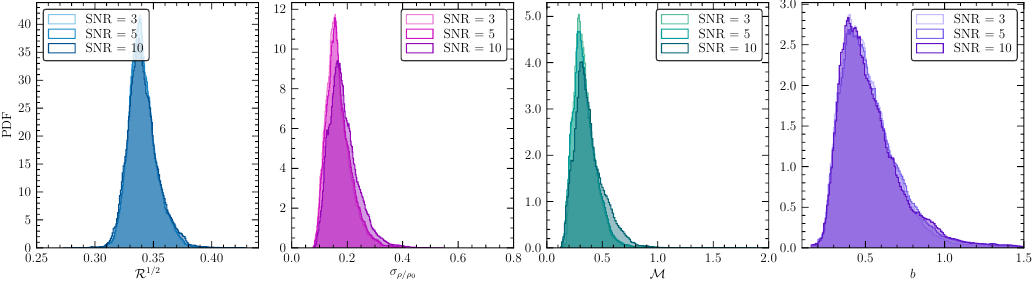}
     \caption{PDFs of the same quantities shown in Fig.~\ref{fig:pdfs}, but for three different signal-to-noise ratios. The data used to construct these PDFs are from inside the contour only. We can see that using a contour of $10\sigma$ provides a robust dataset, which is largely insensitive to the per-channel noise threshold applied to the raw PPV cube. The $\brunt$, $\sigr$, $\mach$, and $b$ values are nearly invariant across the three noise levels we compare here. This gives us confidence that all analysis inside this contour is unaffected by any per-pixel noise.}
     \label{fig:snrPDF}
\end{figure*}
As we previously outlined, we choose a high signal-to-noise ratio (SNR) cut per velocity channel of 10 to ensure the robustness of our data analysis. As defined in Sec.~\ref{sec:obs}, the rms noise level is 1.1~K per 0.98~km\,s$^{-1}$ spectral channel. We apply multiples of this as a per-channel SNR threshold. We investigated three different SNR cuts: 3, 5 and 10. Fig.~\ref{fig:snrPDF} demonstrates that the choice of SNR cut does not impact the analysis quantities particularly, because the data within our analysis contour has a very high SNR. 

\begin{figure*}
     \centering
     \includegraphics[width=\linewidth]{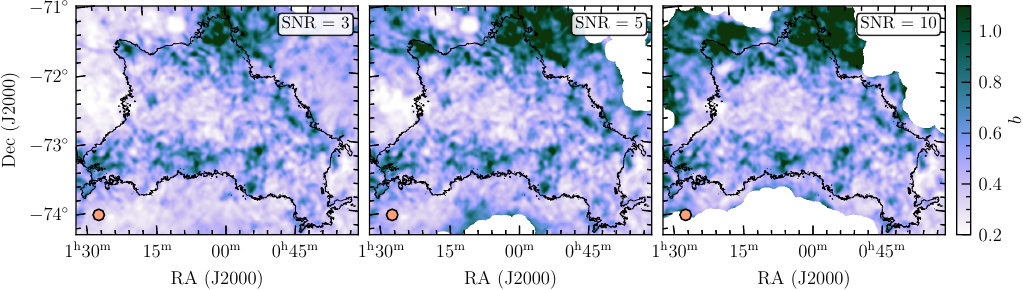}
     \caption{Comparison of the turbulence driving parameter map using different signal-to-noise ratios to threshold the PPV cube prior to running the analysis pipeline. The left panel uses SNR~$=3$, the middle shows SNR~$=5$ and the right panel shows SNR~$=10$ (default used in this study). The orange circle in the bottom left-hand corner of each panel shows the kernel size ($\bpk=10$), and the black contour is the first closed contour as throughout this work.}
     \label{fig:snrcomp}
\end{figure*}
We can see this via visual inspection in Fig.~\ref{fig:snrcomp}, where the nonviable kernels that contain low SNR pixels are shown in white. Clearly, all the kernels inside the analysis contour contain pixels that have velocity channels above the SNR~$=10$ threshold, which accounts for how similar the PDFs of the analysis quantities are. The slight variations are caused by some specific velocity channels in a given pixel being excluded by the per-channel SNR cut, resulting in less channels contributing to the integration that creates the moment-0 and moment-1 maps. This is the case in a minority of pixels towards the contour edge, which in turn result in the slight variations in the PDFs of the analysis quantities. Pixels which contain no velocity channels above the SNR threshold are not considered, and kernels containing such pixels are ignored entirely. This results in the lower SNR thresholds having more viable kernels than the higher SNR thresholds, as shown in Fig.~\ref{fig:snrcomp}.

\section{Correction for optically thick \hi} \label{sec:optical-depth}
\begin{figure*}
     \centering
     \includegraphics[width=\linewidth]{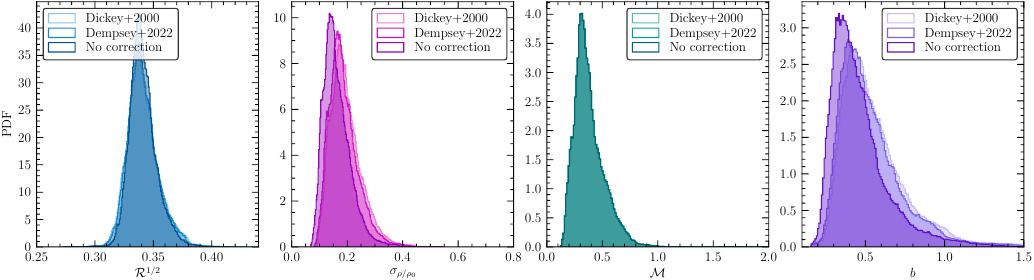}
     \caption{Same as Fig.~\ref{fig:pdfs}, but comparing the influence of correcting the $\hi$ column density for optical-depth effects. We use correction relationships from \citet{Dickey:2000vc} and \citet{Dempsey:2022vi}, and find that $\sigr$ and $b$ increase slightly for both correction cases by \mbox{$12\%$ and $17\%$}, respectively, compared to no HISA correction. The Brunt factor is largely independent of the correction and the Mach number is completely independent of the correction (by definition, as it does not depend on the column density).}
     \label{fig:corrcomp}
\end{figure*}
In this work we chose to account for HISA so as to avoid underestimating the true column density. Using absorption data, the true optical depth of the gas can be estimated and a correction factor derived, such that
\begin{equation}
    \mathcal{R}_{\mathrm{HI}} = \frac{N_{\mathrm{HI,corr}}}{N_{\mathrm{HI,uncorr}}},
\end{equation}
under the assumption that the gas is isothermal. In Fig.~\ref{fig:corrcomp}, we present PDFs of our primary analysis quantities with two different correction factors for the column density applied. The first is from \citet{Dickey:2000vc}, and takes the form,
\begin{equation}
    \mathcal{R}_{\mathrm{HI}} \approx 1+0.667\,(\log_{10}N_{\mathrm{HI,uncorr}} - 21.4),
    \label{eq:dickey}
\end{equation}
such that the correction factor is applied to column densities above $10^{21.4}\,\mathrm{cm}^{-2}$. An updated version of this correction factor was derived by \citet{Dempsey:2022vi} using new ASKAP absorption data, and takes the form,
\begin{equation}
    \mathcal{R}_{\mathrm{HI}} \approx 1+0.51\,(\log_{10}\mathrm{N}_{\mathrm{HI,uncorr}} -21.43).
    \label{eq:dempsey}
\end{equation}
Despite the fact that the \citet{Dempsey:2022vi} correction is only fitted to absorption data in the Bar of the SMC, we chose to apply it to our column density map nonetheless. While the correction factor in the Bar is not directly applicable to data from the Wing (and these are the two main regions enclosed by our analysis contour), a Wing-only correction is not available. By inspecting the data presented in Fig.~10 of \citet{Dempsey:2022vi}, it seems likely that the fit parameters in such a correction would not be too dissimilar to those in the Bar correction, as there is a substantial amount of overlap between the data from the two regions.

Fig.~\ref{fig:corrcomp} shows a comparison of the \citet{Dickey:2000vc}, \citet{Dempsey:2022vi}, and no HISA correction, on the 4~main quantities studied in this work. Without any HISA correction, we find $b = 0.42^{+0.11}_{-0.19}$, but with the \citet{Dickey:2000vc} correction factor we find $b = 0.47^{+0.41}_{-0.16}$ and with the \citet{Dempsey:2022vi} correction we find $b = 0.49^{+0.22}_{-0.13}$. The Mach number is not affected by the correction factor because it is not a function of the column density, but interestingly, the Brunt factor also remains largely invariant, suggesting that the ratio of the column density dispersion to the volume density dispersion is not particularly sensitive to corrections of the $\hi$ intensity.

\begin{figure*}
     \centering
     \includegraphics[width=\linewidth]{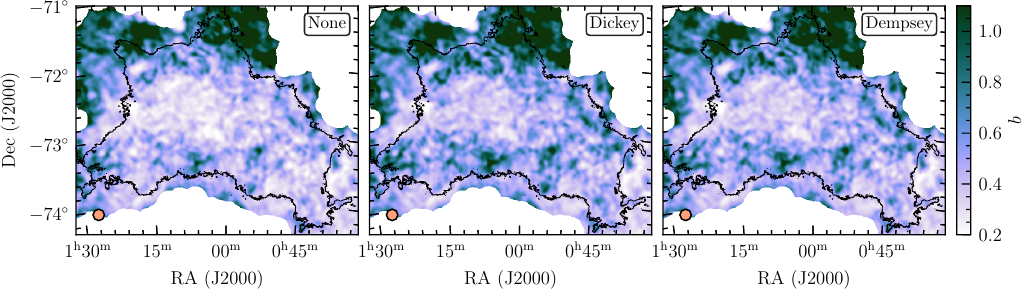}
     \caption{Same as Fig.~\ref{fig:snrcomp}, but for different opacity corrections. The left-hand panel has no correction factor applied. The middle and right-hand panels present maps using the \citet{Dickey:2000vc} (Eq.~\ref{eq:dickey}) and \citet{Dempsey:2022vi} (Eq.~\ref{eq:dempsey}) corrections, respectively.}
     \label{fig:corrcompb}
\end{figure*}
Fig.~\ref{fig:corrcompb} shows the respective maps of $b$ without HISA correction (left), and using the \citet{Dickey:2000vc} (Eq.~\ref{eq:dickey}, middle) and \citet{Dempsey:2022vi} (Eq.~\ref{eq:dempsey}, right) corrections, respectively.

\bsp	
\label{lastpage}
\end{document}